\documentstyle[12pt]{article}

\input{psbox.tex}

\normalsize
\def\sp{~~~~~}

\def\a{\alpha}
\def\b{\beta}

\def\d{\delta}
\def\e{\epsilon}
\def\f{\phi}

\def\h{\eta}

\def\j{\psi}

\def\l{\lambda}
\def\m{\mu}
\def\n{\nu}
\def\o{\omega}
\def\p{\pi}
\def\q{\theta}
\def\r{\rho}
\def\s{\sigma}

\def\x{\xi}

\def\D{\Delta}
\def\F{\Phi}

\def\L{\Lambda}
\def\O{\Omega}

\def\S{\Sigma}

\def\ca{{\cal A}}
\def\cb{{\cal B}}
\def\cc{{\cal C}}

\def\cg{{\cal G}}
\def\ch{{\cal H}}

\def\cl{{\cal L}}

\def\cw{{\cal W}}

\def\rt{\rightarrow}

\def\pa{\partial}

\def\bar#1{\overline{#1}}
\def\Tilde#1{\widetilde {#1}}
\def\Hat#1{\rlap{\kern.10em$\widehat{\phantom G}$}#1}
\def\HAt#1{\rlap{\kern.05em$\widehat{\phantom G}$}#1}

\def\cap#1{\rlap{\kern.1em$\widehat{\phantom{G\vrule height.8em}}$}#1{}}
\def\Cap#1{\rlap{\kern.05em$\widehat{\phantom{G\vrule height.8em}}$}#1{}}

\let\oldtheequation=\theequation
\def\doteqs#1{\setcounter{equation}{0}
	    \def\theequation{{#1}.\oldtheequation}}
\newcounter{sxn}
\def\sx#1{\addtocounter{sxn}{1} \bigskip\medskip \goodbreak
\noindent{\large\bf\centerline{\thesxn.~~#1}} \nobreak \medskip}
\def\sxn#1{\sx{#1} \doteqs{\thesxn}}

\newcounter{axn}

\def\br{\begin{thebibliography}{abc}}
\def\er{\end{thebibliography}}
\def\rf{\bibitem}
\date{}

\begin{document}
\bibliographystyle{unsrt}
\footskip 1.0cm
\thispagestyle{empty}
\setcounter{page}{0}
\begin{flushright}
SU-4240-516\\
April 1993\\
\end{flushright}
\vspace{10mm}

\centerline {\LARGE VERTEX OPERATORS FOR THE $BF$ SYSTEM}
\vspace{5mm}
\centerline {\LARGE AND ITS SPIN-STATISTICS THEOREMS}
\vspace*{15mm}
\centerline {\large A.P. Balachandran \small and \large P. Teotonio-Sobrinho}
\vspace*{5mm}
\centerline {\it Department of Physics, Syracuse University,}
\centerline {\it Syracuse, NY 13244-1130}
\vspace*{25mm}
\normalsize
\centerline {\bf Abstract}
\vspace*{5mm}

Let $B$ and $F=\frac 12F_{\mu \nu}dx^\mu \wedge dx^\nu $ be two forms,
$F_{\mu \nu}$ being the field strength of an abelian connection $A$. The
topological $BF$ system is  given by the integral of $B\wedge F$. With
"kinetic energy'' terms added for $B$ and $A$, it generates a mass for $A$
thereby suggesting an alternative to the Higgs mechanism, and also gives the
London equations. The $BF$ action, being the large length and time scale
limit of this augmented action, is thus of physical interest. In earlier work,
it has been studied on spatial manifold $\Sigma $ with boundaries
$\partial \Sigma $, and the existence of edge states localised at
$\partial \Sigma $ has been established. They are analogous to the conformal
family of edge states to be found in a Chern-Simons theory in a disc. Here
we introduce charges and vortices (thin flux tubes) as sources in the $BF$
system  and show that they acquire an infinite number of spin excitations due
to renormalization, just as a charge coupled to a Chern-Simons potential
acquires a conformal family of spin excitations. For a vortex, these spins
are transverse and attached to each of its points, so that it resembles a
ribbon. Vertex operators for the creatin of these sources are constructed and
interpreted in terms of a Wilson integral involving $A$ and a similar
integral involving $B$. The standard spin-statistics theorem is proved for
this sources. A new spin-statistics theorem, showing the equality of the
``interchange'' of two identical vortex loops and $2\pi $ rotation of the
transverse spins of a constituent vortex, is established. Aharonov-Bohm
interactions of charges and vortices are studied. The existence of
topologically nontrivial vortex spins is pointed out and their vertex
operators are also discussed.

\newpage

\baselineskip=24pt
\setcounter{page}{1}
\newcommand{\be}{\begin{equation}}
\newcommand{\ee}{\end{equation}}

\sxn{INTRODUCTION}\label{sec-introduction}

When a system coupled to an abelian gauge field $A=A_\m dx^\m$ has its U(1)
symmetry spontaneously destroyed, its phenomenology at large length and time
scales is well described by London's constitutive equation
$$dJ=\lambda dA\, , $$
\be
\lambda ={\rm constant}\,, \;\;\;\;\; J:=J_\m dx^\m \label{1.1}
\ee
involving the current $J_\m$.  The conventional Lagrangian description of
(\ref{1.1}) is based on the Nambu-Goldstone field $e^{i\f}$  of unit modulus
and charge $q$.
In $3+1$ dimensions, it reads
$$L_{NG}=\int d^3x\,\cl_{NG} \, ,$$
$$\cl_{NG}= - <H>^2\left( D_\m e^{i\f}\right)^{*}\left( D^\m e^{i\f}
\right) - \frac {1}{4} F_{\m\n} F^{\m\n} \, ,$$
\be
<H>^2=\frac{\lambda }{8q^2}\,, \;\;\;\;\; D_\m e^{i\f} =
(\pa_\m-2\imath qA_\m)e^{i\f} \,  , \label{1.2}
\ee
the metric being $(-1,1,1,1)_{\rm diagonal}$.

It has been known for some time that there is an alternative Lagrangian
approach to London's equation employing the two-form field $B$ instead of
$e^{i\f}$. It is
best explained by first remarking that in the Nambu-Goldstone approach, which
uses (\ref{1.1}), the current
\be
J_\m = 4\imath q<H>^2e^{i\f}D_\m e^{i\f} \label{1.3}
\ee

\noindent solves the constitutive equation (\ref{1.1}) as an identity,
whereas the
continuity equation
\be
\pa_\m J^\m = 0 \label{1.4}
\ee

\noindent is obtained as a field equation.  In contrast, in the alternative
approach, we solve the continuity equation (\ref{1.4}) as an identity by
setting
\be
J^\m = -\frac 12\e^{\m\n\l\r} \pa_\n B_{\l\r}  \, , \label{1.5}
\ee

\noindent the Levi-Civita symbol being fixed by the convention $\e^{0123}=+1$.
The constitutive equation is then obtained as a field equation from the
Lagrangian
$$L_0 = \int d^3x \cl_0  \,  ,$$
$$
\cl _0= \frac 14 \e^{\m\n\l\r}B_{\m\n}F_{\l\r}-\frac {1}{12\lambda }H^{\m\n\r}
H_{\m\n\r} -\frac 14F^{\m\n}F_{\m\n}\,,
$$
\be
H_{\m\n\r}=\pa _\m B_{\n\r} +\pa _\n B_{\r\m} + \pa _\r B_{\m\n}\,.\label{1.6}
\ee

There has been a certain past interest in (\ref{1.6}) when the spatial manifold
has no boundary \cite{gennes,bal,aneziris,bowick}. In previous work
\cite{we,we2}, we also initiated its study
for manifolds $\S$ with boundary $\pa \S$ and established the existence of
edge states localized at $\pa \S$.  They are mathematically analogous to the
quantum Hall edge states \cite{ajit,bimonte,bimonte2}. We also argued
\cite{we,we2} that they are the modes of either of the Lagrangians
\be
L_{\f,\j} = \int_{\pa\S} \dot{\f} d\j \, , \label{1.7}
\ee
\be
L_\f = \frac{1}{2} \int_{\pa\S}\m (\dot {\f}^2 - \o_0 \f^2) \label{1.8}
\ee
on $\pa \S,\j$ being a one form and $\mu $ a volume form on $\pa \S$, and
clarified the relation of (\ref{1.7}) to a coadjoint orbit of a certain group
and its symplectic form \cite{we2}.
$L_{\f, \j}$ is invariant under all the diffeomorphisms (diffeos) of $\S$
whereas
$L_\f$ is invariant under its subgroup $SDiff(\pa\S)$ of diffeos preserving
the
form $\m$.  The Lie algebra of this latter group is related to the algebra
$w_{1+\infty}$  \cite{pope} if $\S$ is the solid cylinder and $\m$ is $d\q dx$
for the
choice of coordinates $(e^{i\q}\e S^1,x\e {\bf R}^1)$ on $\pa \S$. [The wedge
symbols between differential forms will be omitted.]  A generalized Sugawara
construction \cite{goddard} was described.  It was also pointed out that the
edge states were insensitive to excitations in its interior $\S^0$ of $\S$
and that their effective $3+1$ dimensional Lagrangian associated with
(\ref{1.6}) is the $BF$ Lagrangian
$$L^*_0 = \int d^3x \cl^*_0 \, ,$$
\be
\cl^*_0 = \frac 14 \e^{\m\n\l\r}B_{\m\n}F_{\l\r}\label{1.9}
\ee

\noindent when the energy density in $\S^0$ is zero.

In this paper, we explore the $BF$ system with sources calling upon the
experience from our earlier work \cite{bimonte,bimonte2} on the Chern-Simons
Lagrangian.  Two
natural sources for the $BF$ system are magnetic vortices and point charges,
the interaction Lagrangian being \cite{kauffman}
$$
L_I = \int d^3x \cl_I$$
\be
\cl_I = -\frac \l 2\int d\s^1\delta ^3(x-y)\e ^{ab}B_{\m\n}(y)
\pa _ay^\m \pa _by^\n - e\delta ^3(x-y)A_i(z)\pa _0z^i\,, \;\;\;\;\;
\pa _a \equiv \frac {\pa }{\pa \s ^a} \,\,.\label{1.10}
\ee
Here, $\lambda $ and $q$ are constants, $\epsilon ^{ab}=-\epsilon ^{ba}$
with $\epsilon ^{01}=+1$, $i$ is a spatial index taking values 1,2,3,
\begin{eqnarray}
y  : & {\bf R}^1 \times S^1 &\rt {\bf R}^1 \times \S \,,\nonumber \\
     & (\s^0,~\s^1) &\rt y(\s^0,~\s^1)\equiv y(\s) \label {1.11}
\end{eqnarray}

\noindent is the vortex and
\begin{eqnarray}
z & : & { \bf R}^1 \rt {\bf R}^1 \times \S \,  , \nonumber\\
  &   & \s^0 \rt z(\s^0)
\label{1.12}
\end{eqnarray}

\noindent is the charge in spacetime. [The ${\bf R}^1$ factors account for
time, and $\sigma^1$ and $\s^1 +2\p$ are to be identified, in these
expressions. We also assume that $y^0(\s)$, $z^0(\s^0)$ and $\s^0$ are all
equal to coordinate time $x^0$.]
We can also contemplate charged magnetic vortices \cite{aneziris} and these
too will be encountered in the course of our discussion.  We will furthermore
assume that the vortices are unlinked unknots.

In Section 2, we argue that the Lagrangian
\be
L^* = L^*_0 +L_I \label{1.13}
\ee

\noindent requires regularization already at the classical level (that being
also the case with Chern-Simons sources \cite{bimonte,bimonte2}).  The
regularization consists of
first enclosing the location of the vortex or the charge in the interior of a
solid torus or a ball $H$ and letting $H$ shrink to these locations when all
calculations are done.  [The symbol $H$ here is to be thought of as the first
letter of the word ``hole''.]  The presence of this $H$ means that $L^*$ is not
defined on $\S$, but rather on $\S \setminus H$, which is $\S$ with a hole.  In
this way,
it gets associated with a new spatial slice $\S\setminus H$ with a new boundary
$\pa H$
which is a torus $T^2$ or a sphere $S^2$.  This boundary too now acquires edge
excitations exactly as $\pa\S$ does so that the available independent internal
states for a source with a given geographical location are infinite in number.
They are similar to the conformal family of internal states of a Chern-Simons
source.  They are spin  excitations of the sources and of an uncommon
transverse sort for vortices as will be seen in this Section and further
elucidated in Section 3.

In Section 3, we discuss the observables of $L^*$ in detail.  There are first
the class of observables localized at $\pa H$ whose study was already initiated
in the last Section.  They are similar to the observables localized at $\pa
\S$ investigated before \cite{we}.  In addition, there also certain new
observables.
They describe the charges of the sources and at $\pa \S$ (the total charge
adding
up to zero), the magnetic fluxes on the vortices, and their conjugate
variables.
The latter incidentally had turned up at $\pa \S$ before \cite{we} when $\S$
was the solid torus ${\bf T}_{3}$.  The mode decomposition of these
observables is also carried out in this Section.

Section 4 turns to the quantization of the preceding system and introduces
vertex operators.  The conjugate variables alluded to above lead to vertex
operators for $L^*$  just as the variable conjugate to charge enables us to
construct vertex operators in Chern-Simons dynamics.  They are the creation
operators of charges and vortices.  The work also shows the interpretation of
the Wilson integral as a vertex operator for the creation of a charge, and an
analogous expression involving $B$ as the creation operator of a vortex, after
suitable regularization.  This interpretation is the generalization of a
similar interpretation \cite{bimonte2} of the Wilson integral in Chern-Simons
theory.

In Section 5, we prove the spin-statistics theorem.  The conventional
spin-statistics theorem is quickly shown.  We define an operator for an
``interchange'' \cite{ajit2} of two vortices which is a combination of an
exchange and a ``slide" \cite{aneziris,....,liu}.  A new spin-statistics
theorem is
then established, showing the homotopy equivalence of interchange and the
$2\p$ rotation of transverse spin.  Its proof is similar to the proof of the
spin-statistics theorem for
Chern-Simons sources established in \cite{bimonte2}.   The corresponding
quantum operators are then equal for quantizations using covering spaces
\cite{bal2}.
The Section concludes with a discussion of phase changes of states when charges
or charged vortices are transported in loops enclosing fluxes of charged
vortices, or equivalently, strands of vortices. These phase changes may be
thought of as describing ``Aharonov-Bohm" interactions of charges and vortices
or of charged vortices. Quantization conditions involving charges and fluxes
which
make this interaction vanish are derived. The spin-statistics theorems are also
associated with transports of states, and for this reason, the insertion of
these
remarks on the Aharonov-Bohm interaction of charges and vortices in this
Section
seems appropriate.

In this paper, until this point, we largely limit our work to the Lagrangian
$L^*$ which omits the ``kinetic energy" terms $L_0-L_0^*$ for the fields [which
are
proportional to integrals of $\pa_{[\m} B_{\l \r]}\pa^{[\m} B^{\l \r]} {\rm
{}~and}~
\pa_{[\m} A_{\n]} \pa^{[\m} A^{\n]}$. In ref.\cite{we} , in contrast,
we included these terms from the start.
This difference between the papers is not accidental.
The inclusion of $L_0-L^*_0$ does not generate striking differences in the
nature of gauge transformations or the structure of constraints, and these will
in fact be our central pursuits until Section 5.  Edge states of sources and
their vertex operators can in consequence be formally treated just as
previously.  The trouble lies elsewhere:  Sources have divergent self energies
like in electrodynamics, and they require renormalisation, as we will indicate
in Section 6.

Incidentally, by working along parallel lines, it is straightforward also to
generalize the Chern-Simons work of \cite{bimonte} and \cite{bimonte2} to the
Lagrangian.
$$L_{CS}= \int d^2x \cl_{CS} \, ,$$
\be
\cl_{CS} =-\frac 14  (\pa_\m A_\n - \pa_\n A_\m) (\pa^\m A^\n- \pa^\n A^\m) +
\frac {k}{4\p} \e^{\m\n\l} A_\m \pa_\n A_\l  \label{1.14}
\ee

\noindent and its nonabelian version.  (Here the metric has the signature
$(-,+,
+)$ and
$\e^{\m\n\l}$ is the Levi-Civita symbol with $\e^{012}=1$.) This extension has
been carried out in unpublished work \cite{unpublished} elsewhere and is
affected by self-energy divergences just as in the work here.

We have not included kinetic energy terms for sources in this paper as they
do not affect our considerations.

Section 7 is the final one. It discusses certain topologically nontrivial
configurations of transverse spins (``twisted" transverse spins) on vortices
and
their associated vertex operators.

As the final remark of this Section, we note that the conventions regarding
certain factors adopted in this paper differ from \cite{we} and that some
minor algebraic errors of \cite{we} have also been corrected here.

\sxn{How the Source Acquires Spin}\label{sec-source}

The source acquires spin (``transverse" for the vortex) because of
renormalization just like to Chern-Simons source \cite{bimonte2}.  We will
now show how this
happens in detail.  Before we do so, let us record the nonzero Poisson brackets
involving  $A$ and $B$ following from (\ref{1.9}):
\be
\{A_i(x),B_{jk}(y)\}= \e_{ijk} \d^3(x-y) \, .\label{2.1}
\ee

Here, and in what follows,  $\e_{ijk}$  is the Levi-Civita symbol with
$\e_{123}=1$ while PB's, and in fact all considerations, are at equal
times.

\vspace{7mm}

\noindent {\bf 2.1 The Isolated Charge}

The Lagrangian $L^*_0$ has the following two equations analogous to Gauss's law
in
electrodynamics:
\be
\e^{ijk}\pa _iB_{jk}= 0\,,\label{2.2}
\ee
\be
\e^{ijk}F_{jk}= 0 \,. \label{2.3}
\ee
They are obtained by varying $A_0$ and $B_{i0}$ respectively.

In the presence of an isolated point charge at $z=(\vec{z},z^0)$ [$z^0$ being
time $x^0$], the law (\ref{2.2}) gets changed to
\be
\frac 12\e^{ijk}\pa_i B_{jk}(x)=e\d^3(x-z) \label{2.4}
\ee

\noindent as shown by (\ref{1.13}). [All equations are at the equal time
$x^0$ in accordance with a previous remark. We suppress the argument
$\sigma ^0$ in $\vec z(\sigma ^0)$ hereafter.]
The law (\ref{2.3}) is unaffected, if as we for the moment suppose, there are
no
vortices present.

Let $\cb_3$ be a ball or a hole in $\S$ at time $t$ enclosing $\vec{z}$, with
boundary $\pa \cb_3$.  [Earlier, we called $\cb_3$ and $\pa \cb_3$ as $H$ and
$\pa H$ when dealing generically with a charge or a vortex.] According to
(\ref{2.4}), it is then the case that
\be
\int _{\pa \cb_{3}}B =e\, , ~~~~~~~~
 B= \frac {B_{jk}}{2}dx^jdx^k  \label{2.5}
\ee

\noindent where $\pa \cb_3$ is positively oriented. [The orientation of
$\pa {\cal B}_3$ is
inherited from the orientation of the ambient three manifold which is a priori
chosen. The wedge symbols between differential forms are being omitted in this
paper.] On shrinking $\cb_3$ to the point
$\vec {z}$, (\ref{2.5}) shows that $B(x)$ has no definite limit as
$\vec {x} \rt \vec{z}$.  We must thus regularize.

A good way to regularize is to keep $\cb_3$  of finite but small size till all
calculations are done, $\vec{z}$ being in its interior.  $\cb_3$ is
shrunk to
$\vec{z}$  only at the end of the calculations.  [Such a limiting procedure on
holes is hereafter to be understood whenever required.]  When $\cb_3$ is of
nonzero size, the spatial manifold is not $\S$, but $\S \setminus\cb_3$ which
is $\S$ with a hole.  There is then a new boundary $\pa \cb_3$.  According to
\cite{we} (see also \cite{bimonte,bimonte2}), the Gauss laws are then
\be
\cg_0(\l^{(0)}) = \int_{\S \setminus \cb_{3}} \l^{(0)}dB \approx 0 \,,
\label{2.6}
\ee
\be
\cg_1(\l^{(1)}) = 2 \int_{\S \setminus \cb_{3}} \l^{(1)} d A \approx 0, ~~
\l^{(1)} = \l^{(1)}_{j} d x^j,~~
A=A_j dx^j  \label {2.7}
\ee
where the test functions $\l^{(0)}$ and the test 1-forms $\l^{(1)}$ vanish on
boundaries $\pa \S$ and $\pa \cb_3$. We have also introduced the symbol
$\approx$
to denote Dirac's weak equality \cite{dirac}.  If additional boundaries appear
in the
problem, as they will later, $\l^{(j)}$ must vanish there too.

The canonical transformation generated by $\cg_i (\l^{(i)})$ are given by
the PB's
\be
\{\cg_0(\l^{(0)}), A_i (x) \}=\pa _i\l ^{(0)}~~;
\{\cg_0(\l^{(0)}, B_{jk}(x)\}=0 \, , \label{2.8}
\ee
\be
\{\cg_1(\l^{(1)}), A_i(x)\}=0, \{\cg_1(\l^{(1)}),B_{jk}(x)\}=
2(\pa_j\l^{(1)}_k-\pa_k \l^{(1)}_{j}(x))\,. \label{2.9}
\ee

Again, acording to \cite{we,bimonte,bimonte2}, the observables localised at
$\pa \cb _3$ are
\be
q(d \xi^{(0)}) = \int_{\S \setminus\cb_{3}} d \xi^{(0)} B \,  , \label {2.10}
\ee
\be
p(d\xi^{(1)}) = -\int_{\S \setminus \cb_{3}} d \xi^{(1)} A \, , \label{2.11}
\ee

\noindent where $\xi^{(j)}$ are $j$ forms vanishing on $\pa \S$ :
\be
\xi^{(j)}|_{\pa \S} = 0,~~\xi^{(j)}|_{\pa \S} \equiv
{}~{\rm Pull~ back ~of}~ \xi^{(j)}~ {\rm to}~ \pa \S \, . \label {2.12}
\ee
[There are also nonlocal observables to be considered later. When there are
more
boundaries than $\pa \S$ and $\pa \cb_3, ~\xi^{(j)}$ vanish on all boundaries
except $\pa \cb_3$.]
They are observables because their Poisson brackets (PB's) with $\cg_j$ are
weakly zero.  Also all observables with the same boundary values
for $\xi^{(j)}$ are weakly equal and can be identified:
$$
q(d \xi^{(0)}) \approx q (d\tilde{\xi}^{(0)})~~~{\rm if}~~~
\left(\xi^{(0)}-\tilde{\xi}^{(0)}\right)|_{\pa \cb_{3}~{\rm and}~ \pa \S}=
0\, ,
$$
\be
p(d\xi^{(1)}) \approx p(d\tilde{\xi}^{(1)})~~~{\rm if}~~~
\left(\xi^{(1)} - \tilde{\xi}^{(1)}\right)
|_{\pa \cb_{3}~{\rm and}~ \pa \S} =0\, . \label{2.13}
\ee

\noindent This is because their difference becomes a Gauss law on partial
integration.

The new degrees of freedom localized at $\pa \cb_{3}$ are $q$ and $p$.  It is
easy to see that they can be regarded as spin excitations.  Let $R$ be a
diffeomorphism (diffeo) of $\S \setminus \cb_3$ which acts as a rotation of
$\pa \cb_3$ [for the choice  of a round (rotationally invariant) metric on
 $\pa \cb_3=S^2$]
and becomes the identity on $\pa \S$. Let $B \rt R^* B$ and $A \rt R^* A$ be
its
standard actions on the forms $B$ and $A$ by pull back.  The response of the
$q,p$ observables to $R$ is, then,
$$
R:q(d \xi^0) \rt (Rq)(d\xi^{(0)}) =
\int_{\S\setminus \cb_{3}} d\xi^{(0)} R^* B = q(R^{-1*}d\xi^{(0)})\,  ,
$$
\be
R:p(d\xi^{(1)}) \rt (R p) (d\xi^{(1)}) = \int_{\S \setminus\cb_{3}}
d \xi^{(1)}R^*A= p (R^{-1*} d\xi^{(1)}) \,  .\label{2.14}
\ee
Hence $q$ and $p$ change under rotations and carry spin localized at
$\pa \cb_3$.
They describe not just spin of course, as they can be transformed in a similar
way by any diffeo.

There is another way to view these observables.  We can argue that
renormalization has associated a direction, or equally well a point on $S^2=\pa
\cb_3$ to the charge, $\xi^{(j)}$ being fields on these directions or on this
$S^2$.  The particle has thus got ``framed,'' or more precisely acquired a
direction, as a ``spin" degree of freedom.  The spin variables of our source
are not quite this however, being fields $\xi^{(j)}$ on these directions.  The
quantum source associated with $L^*$ is thus described by a first quantized
position and a second quantized direction.

\vspace{7mm}

\noindent {\bf 2.2 The Isolated Vortex}

Next we suppose that there is an isolated vortex or a closed string $y$.
[ Here too, of course,
all considerations are carried out at some fixed time $y^0(\sigma )=x^0$. We
will suppress the argument $\sigma $ in $\vec y(\sigma )$ hereafter.]

In the presence of
the vortex, the law (\ref{2.2}) is not affected, whereas (\ref{2.3}) is
changed to
\be
F_{ij}=\l \e_{ijk}\int d\sigma ^1\partial _1y^k\delta ^3(x-y)\,.
\label{2.15}
\ee

Let $\cc$ be a closed positively oriented contour around the vortex as in
Fig.1.
{}From (\ref{2.15}), we see that
\be
\int_{\cc}A= \l \label{2.16}
\ee
It is thus the case that the vortex is a magnetic line with flux= $\l $.

The orientation of $\cc$ in (\ref{2.16}) is obtained from that of the surface
$S$ cutting the vortex it encloses.  [It cuts it just once as in Figure 1.]  As
for the orientation of the latter, we can fix it as follows.  We first choose
an orientation $\e$ in the ambient three manifold.  Now the orientation $\e_S$
of $S$ along with the direction  of $\pa_1\vec{y}$ where the vortex
intersects $S$ also defines an orientation of the three manifold.  $\e_S$ is
then determined by requiring that the latter is the same as $\e$.

In (\ref{2.16}), by shrinking $\cc$ towards the vortex, we learn that $A(x)$
has
no well defined limit as $x$ approaches the vortex $y$.  We must thus again
regularize.

Regularization can be accomplished much as before by enclosing the vortex
$\vec{y}$ in a  solid torus ${\bf T}_3$ as in Figure 2.
${\bf T}_3$ is of
tiny cross section, and is collapsed
to the vortex after all calculations are over.  [Just as for the hole $\cb_3$
for a charge, this limiting procedure on the hole ${\bf T}_3$ for a vortex is
hereafter to be understood whenever required.]

The Lagrangian $L^*$ is now defined on $\S\setminus {\bf T}_3$ with a
%
%
new boundary $\pa {\bf T}_3=$ the two-torus  $T^2$.  Therefore, the
Gauss laws become \cite{we,bimonte,bimonte2}
\be
\cg_0(\l^{(0)})= \int_{\S \setminus {\bf T}_{3}} \l^{(0)} dB \approx 0 \, ,
\label{2.17}
\ee
\be
\cg_1 (\l^{(1)}) = 2 \int_{\S\setminus {\bf T}_{3}} \l^{(1)} d A
\approx 0 \, , \sp {\l}^{(1)} = {\l}^{(1)}_{j} dx^{j} \label{2.18}
\ee
where
\be
\l^{(j)}|_{\pa \S} = \l^{(j)}|_{\pa {\bf T}_{3}}=0 \, . \label{2.19}
\ee

\noindent Here as before, $\l^{(j)}|_{\pa {\bf T}_{3}}$ for example stands for
the pull back of $\l^{(j)}$ to  $\pa {\bf T}_3$.  If additional boundaries are
inserted, $\l^{(j)}$ must vanish there too.

The construction of the observables also follows \cite{bimonte,bimonte2,we}
and the treatment of
the point charge above.  Thus the observables localized at $\pa {\bf T}_3$ are
essentially all given by
\be
q(d \bar {\x}^{(0)})= \int_{\S\setminus {\bf T}_{3}} d \bar{\x}^{(0)} B  \, ,
\label{2.20}
\ee

\be
p(d \bar{\x}^{(1)}) = -\int_{\S\setminus {\bf T}_{3}} d \bar{\x}^{(1)} A
\label {2.21}
\ee

\noindent where the $j$ forms $\bar{\x}^{(j)}$ vanish on $\pa \S$,
\be
\bar{\x}^{(j)}|_{\pa \S} = 0 \label{2.22}
\ee

\noindent and any other boundary (if such exist) except $\pa {\bf T}_3$.
Furthermore, as in (\ref{2.13}), if two $\bar \x^{(j)}$ agree on
$\pa {\bf T}_3$, the corresponding observables are (weakly) the same.

Let us choose coordinates $\q^i$ mod $2 \p$ for the two-torus $T^2=\pa {\bf
T}_3$ with $\q^2$ becoming the angular coordinate $\s^1$ on the vortex when
${\bf T}_3$ shrinks to the vortex [see Fig. 2].
Then these observables respond to  rotation of $\q^i$ just as in (\ref{2.14}).
Of these, the rotation of $\q^1$ taking a point around a circle such as
$\cc$ of Figure 1, is of greatest interest. Such a rotation has naturally
occured in an earlier work \cite{balclinn} and is associated with a
spin-statistics
theorem to be proved later.  We can thus say that (\ref{2.20}),(\ref{2.21})
describe excitations of this transverse spin.

For specific topologies of $\S$, such as when $\S$ is a three sphere $S^3$,
there can be additional observables localized at $\pa {\bf T}_3$. It is
constructed
along the lines of the observable $P$ in \cite{we}.  Generically, there will
also be nonlocal observables.  We will discuss these local and nonlocal
observables in Section 3.

\sxn{\bf Observables and Their Mode Decomposition}\label{sec-decomposition}

\noindent {\bf 3.1. Local and Nonlocal Observables}

In general, in $\S$, we would have several charges and vortices enclosed by
balls $\cb_3$ and solid tori ${\bf T}_3$.  There is also the ball $\S$ with
$S^2$ as boundary.

Let $\ch$ denote the union of the balls and solid tori enclosing charges
and vortices.  The boundary $\pa(\S\setminus \ch)$ of $\S\setminus \ch$ is the
union of the boundaries $\pa \S$ and $\pa \ch$ of $\S$ and $\ch$.

The constraints ${\cal G}_j$ are now
$$
{\cal G}_0(\lambda ^{(0)})=\int _{\Sigma \setminus {\cal H}} \lambda ^{(0)}dB
\approx 0\,,
$$
$$
{\cal G}_1(\lambda ^{(1)})=2\int _{\Sigma \setminus {\cal H}} \lambda ^{(1)}dA
\approx 0\,,
$$
\be
\lambda ^{(j)}=j~{\rm forms~~with}\;\;\lambda ^{(j)}|
_{\partial (\Sigma \setminus {\cal H})}=0\,.\label{3.3a}
\ee

Let $w^{(j)}$
be closed $j$ forms on $\S \setminus \ch$ and consider
\be
q(w^{(1)})=\int_{\S\setminus \ch} w^{(1)} B \, , \label{3.3}
\ee
\be
p(w^{(2)})=-\int_{\S\setminus \ch} w^{(2)} A \, . \label{3.4}
\ee
We can easily verify that their PB's with $\cg_j$ weakly vanish
and therefore that they are observables.  For example, in view of
(\ref{2.1}),
\begin{eqnarray}
\{\cg_0(\l^0),p(w^{(2)})\} & = & -\int_{\S\setminus \ch} w^{(2)} d\l^{(0)}
\nonumber\\
			   & = &  -\int_{\pa (\S\setminus \ch)}w^{(2)}
\l^{(0)}\nonumber \\
			   & = &  0 \, ,\label{3.5}
\end{eqnarray}
as the test forms $\l^{(j)}$ now vanish on all the boundaries
of $\S\setminus \ch$.

Suppose that we replace $w^{(j)}$ by $w^j+dv^{(j-1)}$ where
\be
v^{(j-1)}|_{\pa(\S\setminus \ch)} = 0 \,. \label{3.6}
\ee
It is then the case that
$$q(w^{(1)}+dv^{(0)})=  q(w^{(1)}) - \cg_0(v^{(0)})\approx q(w^{(1)}) \, ,
$$
\be
p(w^{(2)}+dv^{(1)})=  p(w^{(2)}) -\frac 12 \cg_1(v^{(1)})\approx
p(w^{(2)})\, . \label{3.7}
\ee
Hence $w^{(j)}$ and $w^{(j)} + dv^{(j-1)}$ define equivalent
observables.

The condition (\ref{3.6}) can be relaxed somewhat.  Thus since
$d\hat{v}^{(j-1)} =
d[v^{(j-1)} + u^{(j-1)}]$ if $du^{(j-1)}=0$, we can say that $w^{(j)}$ and
$w^{(j)}+d\hat{v}^{(j-1)}$ define equivalent observables if $\hat{v}^{(j-1)}$
differs from
a $v^{(j-1)}$ fulfilling (\ref{3.6}) by a closed form.
For this purpose, it is sufficient that
$\hat{v}^{(j-1)}|_{\partial (\S \setminus \ch)}$ is exact.  In that case,
$\hat{v}^{(j-1)}|_{\partial (\S \setminus \ch)} = d \hat{w}^{(j-2)}|_
{\partial (\S \setminus \ch)}$.
The form $\hat{w}^{(j-2)}|_{\partial (\S \setminus \ch)}$ can be extended to a
form
$\hat{w}^{(j-2)}$ to all of $\S \setminus \ch$, in fact in many ways, such
that its
restriction to $\pa (\S \setminus \ch)$ is $\hat{w}^{(j-2)}|_{\pa
(\S \setminus \ch)}$.  We can now set
$v^{(j-1)}=\hat{v}^{(j-1)}- d\hat{w}^{(j-2)}$ to see the result.

Hereafter, we write $\ch=\bigcup _\a \ch_a$ where $\pa \ch_a$ is connected.
This
means that an $\ch_\a$ encloses a single charge or a single vortex.

In what follows, we will find it useful to define our meaning of an observable
localized on  a set $U\subseteq \pa \ch_\a$.  We will say that an observable
is localized on $U\subseteq \pa \ch_\a$ iff, up to weak equivalence, it is a
function of observables of type $q(d\x^{(0)})$ or $p(d\x^{(1)})$ [with
integral representations like (\ref{3.3}) and (\ref{3.4})] where
\be
\xi^{(j)}|_{\partial \ch_\b} (\b\neq \a) = \xi^{(j)}|_{\pa\S}=\xi^{(j)}|_
{\pa \ch_{(\a)}\setminus U} = 0\, . \label{3.8}
\ee
This is a good definition because with this condition,
$q(d\x^{(0)})$ for example is determined by $\x^{(0)}|_U$.

In general, there exist observables not accounted for by those localizable on
contractable open sets $U\subseteq \pa \ch_\a$.
For example, let $\S \setminus \ch$ be a ball with two holes $\ch_1$ and
$\ch_2$.
Then there are closed two forms $w^{(2)}_\a (\a=1 ~{\rm or}~2)$ with their
integrals over $\pa \ch_\a$ and $\pa \S$ being nonzero whereas their integrals
over $\pa \ch_{\b}(\b \neq \a)$ vanish.  There are in fact $w^{(2)}_\a $
with $w^{(2)}_\a|_{\pa \ch_{\b}}=0$  if $\b \neq \a$.  How should we think
about the localization properties of $p(w^{(2)}_\a)$?  We suggest the
following
in this regard.  Consider the observables $q(d\b^{(0)})$ and $p(d\b^{(1)})$ of
the sort examined in the last paragraph with
$\b^{(j)}|_{\ch_{\a}}=\b^{(j)}|_{\pa\S}=0$.  They are thus localized
at  $\partial \ch_\b$.  If an observable has nonzero PB with any of  such
observables,
then we propose that they should not be regarded as having null support at
$\partial \ch_\b$.  Instead, they should be thought of as partly living on
$\partial \ch_\b$ as
well.  If all these PB's instead vanish, as it does for example for
$p(w^{(2)}_\a)$, then it is to be thought of as leading to a superselection
rule for $\partial \ch_\b$-localized observables after quantization.  It could
of course vanish because of the nature of its test functions and the Gauss law
constraints.

We have refrained from asserting that observables having zero PB's with
$q(d\b ^{(0)})$ and $p(d\b^{(1)})$ are localized away from $\partial \ch_\b$.
Such an
assertion seems misleading.  For example, if $\x^{(0)}|_\b$ is a constant, and
it is zero on all boundaries except $\b, ~q(d\x^{(0)})$ has the mentioned
feature, and  defines the charge operator for $\partial \ch_\b$, but can not
be reasonably claimed to be localized away from $\partial \ch_\b$.

The preceding paragraphs show that we have to consider two kinds of
observables, namely local observables and those which are not local.  They will
be examined in turn below.

\noindent{\bf 3.2. Local Observables and Their Mode Decomposition}

We recall that two observables are weakly equal if their difference is one of
the constraints in (\ref{3.3a}).  Let $<x>$ denote the equivalence
class of observables weakly equal to $x$.  Hereafter we will call $<x>$ also
as the observable and say that $<x>$ is localized on $\pa \ch_\a$  if $x$
is localized on $\pa \ch_\a$.  In this subsection, we perform the Fourier
decomposition of certain basic $<x>$ which are localized on $\pa \ch_\a$.

It is convenient, for all subsequent discussion, to
choose volume
forms $\m$ on spheres $S^2$ and tori $T^2$ which make up the boundary
components of $\Sigma \setminus {\cal H}$. For the former, we
fix polar coordinates $\q,\f$ and set
$$\m=\frac {\Delta }{4\p } d \cos \q d \f\, ,$$
\be
\int_{S^2} \m = \Delta \, . \label{3.1}
\ee

For the latter, we choose as before angular coordinates $\q^i$ and set
$$\m=\frac{\Delta }{4\p^{2}} d \q^1  d\q^2 \, ,$$
\be
\int_{T^2} \m =\D  \,. \label{3.2}
\ee

Now $\m$ defines a Hilbert space $L^2(\m,\pa
\ch_\alpha )$ of functions on $\pa \ch_\a$ with scalar product
\be
(\chi _1,\chi _2) = \int_{\pa \ch_{\a}} \m ~\chi ^*_1\chi _2 \, . \label{3.9}
\ee

Let $e^{(\a)}_n, n\e{\bf Z}$, define an orthonormal basis of smooth functions
for this space [${\bf Z}$ being the set of integers] where
$e^{(\a)}_0=1/\sqrt{\Delta }:$
\be
(e^{(\a)}_n, e_m^{(\a)})= \d_{nm},\,\,\, e^{(\a)}_{0} = \frac {1}
{\sqrt{\Delta }}\,.\label{3.10}
\ee

When $x_\a$ is localized on $\pa\ch_\a$, it is determined, up to weak
equivalence,
by the pull back of its test forms to $\pa \ch_\a$.  Thus $<x_\a>$ is entirely
determined by the pull back of its test forms to $\pa \ch_\a$.  The Fourier
decomposition of $<x_\a>$ is thus accomplished by the Fourier decomposition of
these pull backs.

This Fourier decomposition can now proceed as in \cite{we}.  The observables
$q_n(\a)$ are defined by
\be
q_n(\a) = <q(d\xi^{(0)}_n)>\,,~~~\xi^{(0)}_n|_{\pa \ch_{\a}}=e^{(\a)}_n\,.
\label{3.11}
\ee
[It is understood here that $\x^{(j)}_n$ vanish on all boundaries
except $\pa \ch_\a$.]  The observables $p_n(\a)$ are given by
\be
p_n(\a) = <p(d\xi^{(1)}_n)>,~d\xi^{(1)}_n|_{\pa \ch_{\a}}=e^{(\alpha )*}_n\,
\m,n\neq 0\,.\label{3.12}
\ee

The value $n=0$ can not be excluded from (\ref{3.11}).  For although
$e_0^{(\alpha )}$ is a
constant,  we can not set $\x^{(0)}_0=e_0^{(\alpha )}$ since
$\x^{(0)}_0|_{\ch_{\b}}\,(\b\neq\a)$ and $\x^{(0)}_0|_{\pa\S}$ must vanish.
In contrast, the value $n=0$ is excluded from (\ref{3.12}) because
$e^{(\alpha )}_0
\m = \frac{1}{\sqrt{\Delta }}\m$ is not exact unlike $d\x^{(1)}_0|_{\ch_{\a}}=d
(\x^{(1)}_0|_{\ch_{\a}})$.

The nonzero PB's involving $q^{(\a)}_n$ and $p^{(\b)}_n$ are,  by a natural
definition, given by
\be
\{q_n (\a),p_m (\b)\} :=\d_{\a\b} < \{q(d\xi^{(0)}_n), ~p(d\xi^{(1)}_m)\} >=
\d_{\a\b} \d_{nm}\,. \label{3.13}
\ee

We here record our choices of $e^{(\a)}_n$ for $S^2$ and $T^2$ made in
\cite{we}, choices which we will adopt in this paper as well.
For $S^2$, they are defined by the correspondences
$$
n\rt Jm\,,
$$
\be
e^{(\alpha )}_n\rt e_{JM}=\left (\frac{4\p}{\Delta }\right)^{\frac{1}{2}}
Y_{Jm},~Y_{Jm}~=~{\rm Spherical~ harmonics}~ \,, \label{3.14}
\ee
$e^{(\a)}_0$ becoming $\frac{1}{\sqrt \Delta }$.

For $T^2$, we have the correspondences
$$
n \rt \vec{N} = (N_1,N_2),
$$
$$
e^{(\a)}_n(\vec{\q}) \rt e_{\vec{N}}(\vec{\q}) = \frac{1}{\sqrt{\Delta }}
e^{i\vec{N}\cdot \vec{\q}},
$$
\be
\vec{\q}= \q^1,\q^2,~~~\vec{N}\cdot \vec{\q}=N_1\q^1+N_2\q^2,\,\,~~ N_i \in
{\bf Z}\, . \label{3.15}
\ee

For general topologies, there are more $p$-type observables localized at $\pa
\ch_{\a}$.  This would be the case if $\pa \ch_\a$ admits closed but inexact
one forms $\omega _N^{(1)}|_{\pa \ch_{\a}}$, the one forms $\omega _N^{(1)}$
on $\S\setminus \ch$
vanishing on $\pa (\Sigma \setminus \ch_\b)$.  Here as usual
$\omega ^{(1)}_N|_{\pa \ch_{\a}}$
is the pull back of $\omega ^{(1)}_N$ to $\pa \ch_{\a}$. We can then choose
$\omega ^{(1)}_N$ and cycles $\cc _M\subset \pa \ch_\a$ for the generators of
the
homology group of $\pa \ch_\a$ such that
\be
\int_{\cc_{M}}\omega ^{(1)}_N = \d_{MN} \times ~{\rm nonvanishing ~constant}
{}~\,.\label{3.16}
\ee


It is our experience that $\cc_M$, regarded as a cycle in $\S \setminus
\ch$, is contractible to a point (if there are no links) or in other words
homologous to the trivial cycle
consisting of a point.  That being so, $\omega ^{(1)}_N$ can not as a rule be
closed in
$\S \setminus \ch$, it is only its pull back to $\pa \ch_\a$ which can enjoy
this property.

Referring to (\ref{3.12}), we see that the preceding Fourier decomposition
misses the observables
\be
P_N(\a)=<p(d\o^{(1)}_N)>\, , \label{3.17}
\ee
$d\omega ^{(1)}_N|_{\pa \ch_{\a}}$ being zero.  They must thus be added to
our list of observables.

It is to be noted no new observable is associated with $\x^{(1)}_N$ having
$\x^{(1)}_N|_{\pa \ch_{\a}}$ exact.  This is so for the following reason:
If $\x^{(1)}_N|_{\pa \ch_{\a}}$ is exact, we can write
$\x^{(1)}_N|_{\pa \ch_{\a}}= d \hat{\x}^{(0)}_N|_{\pa \ch_{\a}}$.  Now
$\hat{\x}^{(0)}_N|_{\pa \ch_{\a}}$ can be readily extended to a function
$\hat{\x}^{(0)}_N$ defined in all of $\S \setminus \ch$, and
$p(d \x^{(1)}_N) \approx p(d d \hat{\x}^{(0)}_N)= 0.$ Thus $<p(d
\x^{(1)}_N)>=0$.

An example of an observable of type $P_N(\a)$ was encountered in \cite{we}
where the
space $\S={\bf T}_3$ was considered without holes, and the corresponding
$P_N(\a)$ was called $P$. This space $\S$ is the {\em interior} of the torus
in Fig. 2.  If $\omega ^{(1)}$ is the one form leading to $P$, then
\be
\omega ^{(1)}|_{\pa \S}= -\frac{\sqrt{\Delta }}{~4\p^{2}} d\q^1 \,.
\label{3.18}
\ee

\noindent [$\q^1$ and $\q^2$ in Fig. 2 are to be indentified with $\q^2$ and
$\q^1$ of \cite{we}.]

We can make up an example of this kind whenever there is a cycle $\cc$
in $\pa(\S \setminus \ch)$  which can not be contracted to a point while
staying within $\pa(\S \setminus \ch)$.
Thus $\cc$ here, when regarded as defining an element of the
homology group $H_1(\pa(\S \setminus \ch))$, defines a nontrivial element of
that group.
When
there is such a $\cc$, we can find a closed one form $\omega ^{(1)}|_
{\pa \ch_{\a}}$
with a nonzero integral over $\cc$, and a zero integral over all other cycles
in $\pa\ch_\a$.  This form can also be clearly extended as a one form
$\omega ^{(1)}$ over $\S \setminus \ch$ with
$\omega ^{(1)}|_{\pa \ch_{\b}}=0$ for $\b \neq \a$. [This form $\omega ^
{(1)}$ over
$\S \setminus \ch$ must not be closed in order to get a nonzero observable,
only its pull back to $\pa \ch_\a$ must be
closed.] Such an $\omega ^{(1)}$ gives us an example of the sort we want.

We next note that
\be
\{p_n(\a),~P_N(\b)\}=\{P_N(\a),~P_M(\b)\}=0\, ,\label{3.19}
\ee
\be
\{q_n(\a),~P_N(\b)\}= \d_{\a\b} \int_{\pa \ch_{\a}}
e_n^{(\a)} d\omega ^{(1)}_N = 0 \,. \label{3.20}
\ee

\noindent (\ref{3.19}) here is evident.  Thus $P_N(\a)$ define superselection
rules for
observables at $\pa \ch_\a$.  Their physical meaning will be addressed in
Section 3.3.

$P_N(\a)$ are not the only observables leading to superselection rules.
$q_0(\a)$ also has this property:  its PB with all observables localized at
$\pa \ch_\a$ vanish in view of (\ref{3.13}) and because $m\neq 0$.

The physical meaning of $q_0(\a)$ is that it is a measure of charge contained
by the hole $\ch_\a$ as we shall see in Section 3.3.

In suggesting that $q_0(\a)$ and $P_N(\a)$ are associated with superselection
sectors, we have omitted an examination of their PB's with the nonlocal
observables.  We will see below that there are observables of this kind related
to charge and vortex creation.  In what follows, arguments will be presented
to justify this neglect.

\noindent {\bf 3.3. Nonlocal Observables}

We find by examples that these observables  are associated with charge and
vortex creation, the former being conjugate to $q_0(\a)$ and the latter to
$P_N(\a)$.  It is also the case that the former can be interpreted in terms of
line integrals of $A$ (Wilson line integrals) and the latter in terms of
surface integrals of $B$.

Nonlocal observables, or rather their suitably regularized exponentials [see
Section 4], are the analogues of vertex operators in conformal field theories
(CFT's) \cite{goddard}.  They have this relationship only in a generalized
sense, being
classical, in 3+1 dimensions and in a field theory distinct from a CFT.  But
they do resemble CFT vertex operators in spite of these differences.

Now in CFT, a vertex operator is not treated as an observable.  It is, rather,
an intertwining operator between inequivalent representations of the affine
Lie algebra with distinct charges or momenta.

In a similar spirit, the nonlocal ``observables'' we now describe in their
classical versions are perhaps more properly regarded in quantum theory as
operators intertwining inequivalent representations of the algebra of
the remaining observables, and not as observables themselves.

We tacitly accepted this interpretation when suggesting previously that
$q_0{(\a)}$ and $P_N(\a)$ lead to superselection rules.  Perhaps a physical
argument can be found which explains why nonlocal ``observables'' are not in
fact observables.

Hereafter, we will call nonlocal ``observables'' as nonlocal variables or
nonlocal operators to emphasize that we do not treat them as observables.
\vspace{3mm}

\noindent
{\bf i) The Conjugate of Charge}

It is enough to consider the ball $\S$ to have a single hole $\ch_1=\cb_3$ to
illustrate the ideas behind this variable.  The hole could have been put there
to regularize a charge.  In quantum theory, it could also have been punched as
a preparatory move to create a charged state there using a vertex operator as
in Section 4.  [Such a procedure has been described in \cite{bimonte2} for
Chern-Simons theories.]  For what follows here, $\ch_1$ can also be
${\bf T}_3$  and punched in $\cb_3$ to regularize a vortex.

Now $\S\setminus \ch_1$ admits a closed two form $\omega ^{(2)}$ which is not
exact. A closed $\omega ^{(2)}$ with its pull back to $\ch_1$ being $\m$ is one
such $\omega ^{(2)}$.
If $S^2$ is a two-sphere enclosing $\ch_1, ~\omega ^{(2)}$ is in general
specified by the properties
\be
d\omega ^{(2)}=0\, ,\label{3.21}
\ee

\be
\int_{S^2} \omega ^{(2)} \neq 0 \,. \label{3.22}
\ee

Consider
\be
W(\omega ^{(2)}) = \int_{\S\setminus \ch_{1}} \omega ^{(2)} A \,.\label{3.23}
\ee
Under a gauge transformation $A \rt A+d\l^{(0)}$ where  $\l^{(0)}|
_{\pa(\S\setminus\ch_{1})}=0$, the function $W$ does not change:
\begin{eqnarray}
W(\omega ^{(2)}) & \rt & W(\omega ^{(2)})+ \int_{\S\setminus \ch_{1}}
\omega ^{(2)}d\l^{(0)}\nonumber \\
& = & W(\omega ^{(2)})+ \int_{\pa(\S\setminus \ch_{1})}
      \omega ^{(2)} \l^{(0)}\nonumber \\
& = & W(\omega ^{(2)})\,.\label{3.24}
\end{eqnarray}

\noindent It thus has zero PB with $\cg_{0}(\l^{(0)})$, and evidently with
$\cg_1(\l^{(1)})$ as well, and can be thought of as an observable.
Alternatively,
the observable is the equivalence class $<W(\omega ^{(2)})>$.

Also if $q_0(1)$ is the
charge in $\ch_1$, as we argue under iii) below, we find using (\ref{3.11})
that
\be
\{W(\omega ^{(2)}), q_0{(1)}\} =\int _{\Sigma \setminus \ch }\omega ^{(2)}
d\xi _0^{(0)}=\frac {1}{\sqrt \Delta } \int_{\ch _1} \omega ^{(2)}\, .
\label{3.25}
\ee

\noindent Thus, $W (\omega ^{(2)})$ is conjugate to charge and the charge
creation operator is
a suitable exponential constructed using it. [See Section 4.]

There is an interpretation of $W(\omega ^{(2)})$ in terms of the Wilson
integral as we now indicate.

Let us choose $\omega ^{(2)}= \bar{\omega }^{(2)}$ where $\bar{\omega }^{(2)}$
has support on a  thin
tube $T$ of cross section $\d$ as pictured in Fig.3.  In that figure,
$L$ is a
line in the middle of $T$.  In the limit $\d \rt 0$, we then find,

$$W(\bar{\omega }^{(2)}) \rt \tilde \l \int_{L} A\, ,$$
\be
\tilde \l = \int_{S^{2}} \bar \omega ^{(2)} \,. \label{3.26}
\ee
The Wilson integral is thus a limiting form of $W(\bar{\omega }^{(2)})$.

The Wilson integral for two lines $L$ and $L^\prime $ describe the same
observable provided only that $L'$ can be deformed to $L$
holding $P$ and $Q$ fixed.
Fig. 4 shows such an $L$ and an $L^\prime $. This
is so because the two integrals are seen to differ by a constraint on using
Stokes' theorem.  It follows that
the integral of $A$ over $L$ is associated with a localized blip at
$P$ and, in the limit of $\ch_1$ shrinking to $C$, a direction, shown by an
arrow in Figs. 3 and 4, attached to the particle position.  It is this
directional degree of freedom which leads to the spin excitations of the
particle.

[Section 4 contains further discussion of the dependence of
$<W(\omega ^{(2)})>$ on $L$.]

The variable $<W(\omega ^{(2)})>$
is nonlocal, its associated charges being at the two boundaries $\pa \ch_1$ and
$\pa \S$.

The variable
\be
\langle \int_{L} A \rangle \label{3.27}
\ee

\noindent can be Fourier analyzed. We reserve this task to Section 4.
In the Chern-Simons field theory, a
corresponding analysis is known to lead to the Fubini-Veneziano field and the
associated vertex operator \cite{bimonte2,goddard}.

The PB of $<W(\omega ^{(2)})>$ with the rest of the observables is
sensitive to
the choice of $\omega ^{(2)}|_{\pa \ch_{1}}$.  The choice leading to simple
answers is
\be
\omega ^{(2)}|_{\pa \ch_{1}}  = \m \,. \label{3.28}
\ee
Then
\begin{eqnarray}
\{<W(\omega ^{(2)})>,~q_n(1)\} & = & \int_{(\partial \S\setminus\ch_{1})}
\omega ^{(2)} d \x^{(0)}_{n}
\nonumber \\
		       & = &  \int_{\S\setminus\ch_{1}} \m
e^{(1)}_n = \sqrt{\Delta } \d_{n0} \, ,\label{3.29}
\end{eqnarray}

\be
\{<W(\omega ^{(2)})>,~ p_n (\a)\} = 0 \, . \label{3.30}
\ee

We will have to consider yet another variable, the conjugate flux, in
ii).  The PB involving it will be recorded there.  A summary of all PB's can
be found in iv) below.

Note that (\ref{3.28}) and (\ref{3.29}) are correct even if $\ch_1$ is ${\bf
T}_3$, provided the closed form $\omega ^{(2)}$ satisfies (\ref{3.28}).

For the manifold $\S \setminus\ch_1$ we are considering, a similar analysis
can be done on $\pa \S = S^2$.  The analogues of the observables $q_n(\a)$ and
$p_n (\a)$ on this $S^2$ and their PB's, have been found in \cite{we}.
One can  readily check that
relations like (\ref{3.29}) and (\ref{3.30}) also hold.

For more complicated or for alternative topologies too, such as for
$\S\setminus \ch$ with $\ch$
having several disconnected components $\ch_\a$, or with $\pa \S$ being ${\bf
T}^2$, similar conclusions can be drawn by chosing $\omega ^{(2)}$ to be $\m$
on one $\pa \ch_\a$ and $\pa \S$, and zero on $\pa \ch_\b (\b \neq \a)$.

\noindent {\bf ii) The Conjugate of Flux}

Experience with $W(\omega ^{(2)})$ suggests that this nonlocal variable
involves
closed but inexact one forms $\omega ^{(1)}$, the variable then being the
equivalence class of
\be
V(\omega ^{(1)})= \int \omega ^{(1)}B \, . \label{3.31}
\ee
The integration here is over $\S$, or if it has holes, over
$\S\setminus \ch$.

$V(\omega ^{(1)})$ has vanishing PB with the constraints so that
$<V(\omega ^{(1)})>$ is
first class \cite{dirac}.  We only need to examine the response of
$V(\omega ^{(1)})$ to the
transformation $B \rt B+ d \l^{(1)}$ to verify this fact.  Here $\l^{(1)}$
vanishes on all the boundaries.  Since

\be
\int \omega ^{(1)} d \l^{(1)}= 0, \label{3.32}
\ee

\noindent $\omega ^{(1)}$ being closed, the result follows.

In \cite{we}, an example of $\omega ^{(1)}$ was presented when $\S$ was the
solid torus
interpreted as the {\em interior} of ${\bf T}_3$ in Fig. 2.  The integration
for $V(\omega ^{(1)})$ is then over this $\S$ while
\be
\omega ^{(1)}|_{\pa\S} = d\q^2 \,. \label{3.33}
\ee

It is easy to see that there is a closed $\omega ^{(1)}$ with the property
(\ref{3.33}).  For instance, we can introduce $\q^1,~\q^2$ and a third
variable
$r$ as coordinates for all of ${\bf T}_3,~r$ being a radial distance from the
central dotted thread in Fig.2, with $r=1$ say for $\pa {\bf T}_3$.  The
meaning of $\q^i$ are as shown in that Figure.
We can then set
\be
\omega ^{(1)}=d \q^2 \label{3.34}
\ee
in all of ${\bf T}_3$.

Now if $L$ is as indicated in Fig. 2,
\be
\{ V(\omega ^{(1)}), \int_L A \}=-\int _L\omega ^{(1)}=-2\pi \,.\label {3.35}
\ee

As the integral of $A$ over $L$ measures the flux enclosed by
$L,~V(\omega ^{(1)})$
is conjugate to magnetic flux.  This flux is confined to $\pa {\bf T}_3$  as
$L$ can be anywhere in the interior of ${\bf T}_3$.  The creation operator of a
vortex on $\pa {\bf T}_3$ is thus a suitable exponential constructed from
$V(\omega ^{(1)})$.  [See Section 4.]

Previously, in (\ref{3.26}) $W(\omega ^{(2)})$ was interpreted in terms of an
integral over $A$. There is a similar interpretation of $V(\omega ^{(1)})$ in
terms of
$$\int_{\bar{\S}} B$$

\noindent where $\bar{\S}$ is the cross section of ${\bf T}_3$ for the
constant value $\bar{\q}^2$ of $\q^2$.  We
find this interpretation as follows.   Let $\bar{S}_\d$ be the solid disc
with $\q^2$ in the range $\bar{\q}^2-\frac{\d}{2}<\q^2 <\bar{\q}^2 +
\frac{\d}{2}$ as in Fig.5.
Let us choose $\omega ^{(1)}= \bar \o^{(1)}$ where
$\o^{(1)}$ has support on $\bar S_\d $.
In the limit $\d \rt 0$, we then find,
\begin{eqnarray}
V(\bar \omega ^{(1)}) & = & \L \int_{\bar{S}} B \,,~~~\bar{S}=\bar{S}_0,
\nonumber \\
\L & = & \int_{L} \bar \omega ^{(1)}\,. \label {3.36}
\end{eqnarray}

In this example, we have called $V(\omega ^{(1)})$ nonlocal even though it is
associated with a vortex at a single boundary.  This usage may not be
inappropriate.  Thus the form $\omega ^{(1)}$ is closed but not exact and can
not
therefore be chosen to be zero outside a small neighbourhood of $\pa {\bf
T}_3$.  In contrast,
in (\ref{3.11}) and (\ref{3.12}), the forms $\x^{(j)}_n$ characterizing an
observable localized at $\ch_\a$ can always be made zero outside a small
neighbourhood of $\ch_\a$ without affecting the observable.

Reference \cite{we} can be consulted for further discussion of this example.
It can
be generalized.  For illustrative purposes, we will now indicate a few such
generalizations.
\vspace{5mm}

\noindent ${\bf \S \setminus \ch_{\rm 1} = \cb_{\rm 3} \setminus T_{\rm 3}}$

Let $\S = \cb_3$  and let us dig a single hole $\ch_1= {\bf T}_3$ in $\S$.
This hole is pictured in Fig. 2. It could have been put there to regularize a
vortex. In quantum theory, just as for the charge \cite{bimonte2}, it could
also have
been punched as a preparatory move to create a vortex state using a vertex
operator. [See Section 4.]

Now $\S\setminus \ch_1$ admits a closed but inexact  one form $\omega ^{(1)}$.
Its cohomology class is entirely described by saying that $\omega ^{(1)}|_
{\pa \ch_{1}}=d\q^1$.

Let $\cc$ be a contour enclosing this ${\bf T}_3$ as shown in Fig. 6.  Then
\be
\{V(\omega ^{(1)}), \int_{\cc} A \} = -\int _{\cal C}\omega ^{(1)}=-2\pi  .
\label{3.37}
\ee
Since $\int_{\cc} A$ is a measure of the flux through
$\cc$, and $\cc$ can be as close to $ \pa {\bf
T}_3$ as we please without affecting (\ref{3.37}), this shows that
$V(\omega ^{(1)})$
is conjugate to magnetic flux threading $\ch_1$ in the $\q^2$ direction.  The
creation operator of a vortex  along the line $L$ of Fig. 2 is hence
an exponential involving $V(\omega ^{(1)})$ in the limit that $\ch_1$ shrinks
to $L$.

Let $\bar {\s}$ be a surface spanning from $\pa {\bf T}_3$ to $\pa \S$ as
in Fig. 7.  Just as in (\ref{3.35}), we can then show by a suitable choice of
$\omega ^{(1)}$  and a limiting procedure that
\be
V(\omega ^{(1)}) = 2\p \int_{\bar{\s}} B \label{3.38}
\ee
where we have assumed that $\omega ^{(1)}|_{\pa {\bf T}_{3}}= d \q^1$ and
hence that
\be
\int_{\cc} \omega ^{(1)} = 2\p \,. \label {3.39}
\ee

Just as we argued previously in the case of $<W(\omega ^{(2)})>$, we can argue
here
as well that the distortion of $\bar{\s}$ to another surface $\Tilde{\s}$
as in Fig. 7 without altering its boundaries does not affect
$<V(\omega ^{(1)})>$. What this means is that $V(\omega ^{(1)})$ describes a
vortex at $\pa \bar {\s}\bigcap \pa {\bf T}_3$ and another one at
$\pa \bar {\s}\bigcap \pa \S $, but does
not describe any degree of freedom in the interior of $\S \setminus \ch_1$.
Note that in
the limit where $\ch_1$ shrinks to $L$ (represented in Fig.7 by a dotted line
inside ${\bf T}_3$), the integral of $B$ over $\bar {\s}$
gets associated not merely with $L$, but also with a set of directions
on $L$ pointing from $L$ to $\bar {\s} \bigcap \pa {\bf T}_3$  as shown in
Fig. 7.
This field of directions endows the vortex with a transverse spin degree of
freedom along the lines discussed previously in \cite{balclinn}. Its
direction can be defined by choosing a metric and using it to define the
tangent to $\bar \s$ at a point of the vortex which is normal to its slope
there. Then the transverse spin can be said to point in the direction of
this tangent.

It is possible to ``Fourier analyse'' $<V(\omega ^{(1)})>$ for the
$V(\omega ^{(1)})$ of (\ref{3.38}) just as (\ref{3.27}) can be subjected to
such an analysis. The resultant
field however is no ordinary field, but a string field dependent on
$\pa \bar {\s}\bigcap \pa {\bf T}_3$ and $\pa \bar {\s}\bigcap \pa \S$.
We discuss this field further in Section 4.

The PB of $<V(\omega ^{(1)})>$ with the rest of the observables at a boundary
$\pa \ch_\a$ is sensitive to the choice of $\omega ^{(1)}|_{\pa \ch_{\a}}$.
For $\S \setminus \ch_1$, let us choose
\be
\omega ^{(1)}|_{\pa \ch_{1}}  = d \q^1 \,. \label{3.40}
\ee
We then get
\be
\{<V(\omega ^{(1)})>,~q_{\vec{N}}(\a)\}= 0\, ,\label{3.41}
\ee

\be
\{<V(\omega ^{(1)})>,~p_{\vec{N}}(\a)\}=\int d\theta ^1d\x_{\vec N}^{(1)}=
-\int d\left( d\theta ^1\xi ^1_
{\vec N}\right)=- \int _{\partial {\cal H}_1}d\q^1 \x^{(1)}_{\vec{N}}=0\,.
\label{3.42}
\ee
Here $N_1$ and $N_2$ are not both
zero and $\a$ of course is $1$. Also the conclusion in (\ref{3.42}) follows
from (\ref{3.12}) which permits us to assume that
$\x^{(1)}_{N}|_{\pa \ch_{\a}}=
\frac{e^{-i \vec{N}\cdot \vec{\q}}}{(-iN_{1})}d\q^{2}$ or
$\frac{e^{-i \vec{N}\cdot \vec{\q}}}{(-iN_{2})}d\q^{1}$ according as $N_1$ or
$N_2$ is nonzero.

We have yet to look at $\{<V(\omega ^{(1)})>,~P_N(\a)\}$ and
$\{<W(\omega ^{(2)})>,\,<V(\omega ^{(1)})>\}$. We have
\be
\{<V(\omega ^{(1)})>,\,P_N(\alpha )\}=-\int _{\partial \ch_\a}\omega ^{(1)}
\omega ^{(1)}_N\,,\label{4.43a}
\ee
\be
\{<W(\omega ^{(2)})>,~ <V(\omega ^{(1)})>\} = \int_{\S\setminus \ch_{1}}
\omega ^{(2)}\omega ^{(1)}\label{3.43}
\ee
where $\omega ^{(2)} \omega ^{(1)}$ as usual denotes $\omega ^{(2)}\wedge
\omega^{(1)}$.

We will consider the evaluation of (\ref{3.43}) for two typical cases, namely
$\S=\cb_{3}$ and ${\bf T}_3,~\ch_1$ in both these instances being a solid
torus.  The choice $\S={\bf T}_3$ actually gives us two
examples depending on the placement of $\ch_1$.  They are illustrated
in Figs. 8 and 9.
It is to be noted that Figs.~~8 and 9 admit two independent $\omega ^{(1)}$
corresponding
to the two cycles $\cc_i$.  As for $\omega ^{(2)}$, it can in all
three cases be taken
to be the closed two form which reduces to our canonical volume form when
pulled back to $\pa \S$ and $\partial \ch_1$.

Let $\o^{(j)}_{0}$ be closed $j$ forms with same boundary values as
$\o^{(j)}$,
\be
\o^{(j)}_{0}|_{\pa(\S\setminus\ch_{1})} = \o^{(j)}|_{\pa(\S\setminus
\ch_{1})}, \label{3.44}
\ee
such that $\int_{\S\setminus \ch_{1}} \o^{(2)}_{0} \o^{(1)}_{0}$
can be calculated.  We can write
\begin{eqnarray}
\int_{\S\setminus \ch_{1}} \o^{(2)}\o^{(1)} & = & \int_{\S\setminus \ch_{1}}
\o^{(2)}_{0}\o^{(1)}_{0} + \int_{\S\setminus \ch_{1}} \D \o^{(2)}\o^{(1)}_{0}
 \nonumber \\
 & + & \int_{\S\setminus \ch_{1}} \o^{(2)}_{0}\D \o^{(1)} +
\int_{\S\setminus \ch_{1}}\D \o^{(2)}\D \o^{(1)}\, ,\nonumber\\
\D \o^{(j)} & = & \o^{(j)} - \o^{(j)}_{0}\, .
\label{3.45}
\end{eqnarray}

We now show that the last three terms here and hence (\ref{3.43}), can be
calculated.

We argue as follows to compute these terms. First note that $\D \o^{(j)}$ is
closed and vanishes on the boundaries:
\be
d \D \o^{(j)} = 0 \, ,\label{3.46}
\ee
\be
\D \o^{(j)}|_{\pa (\S \setminus \ch_{1})} = 0 \,. \label{3.47}
\ee

For the topologies we consider, every cycle in $\S\setminus \ch_1$ is
homologous to a cycle on the boundary.  The integral of $\D \o^{(j)}$ over
every
such cycle is zero by (\ref{3.47}), and in view of (\ref{3.46}), its integral
over {\em all} cycles in $\S \setminus \ch_1$ is zero.  So $\D \o^{(j)}$ is
exact:
\be
\D \o^{(j)} = d \e^{(j-1)} \, .  \label{3.48}
\ee

\noindent Here by (\ref{3.47}), $\e^{(j-1)}|_{\pa (\S \setminus \ch_{1})}$ is
closed:
\be
d \e^{(j-1)}|_{\pa (\S\setminus \ch_{1})} = 0\, . \label{3.49}
\ee

Using these results, we find,
\be
\int_{\S\setminus \ch_{1}}\D \o^{(2)} \o^{(1)}_{0} =
\int_{\pa(\S\setminus \ch_{1})}\e^{(1)} \o^{(1)}_{0} \, , \label{3.50}
\ee
\be
\int_{\S\setminus \ch_{1}} \o^{(2)}_{0}\D \o^{(1)}=
\int_{\pa(\S\setminus \ch_{1})} \o^{(2)}_{0}\e^{(0)}\, , \label {3.51}
\ee
\be
\int_{\S\setminus \ch_{1}}\D \o^{(2)}\D \o^{(1)} =
\int_{\S\setminus \ch_{1}}d \e^{(1)}d\e^{(0)} =
\int_{\pa(\S\setminus \ch_{1})}\e^{(1)}d \e^{(0)}= 0 \, . \label{3.52}
\ee

There are two sorts of connected boundaries, namely $S^2$ and $T^2.~S^2$ occurs
as $\pa \S$ for $\S=\cb_3$ whereas $T^2$ is $\pa \ch_1$ and also $\pa \S$~ if~
$\S={\bf T}_3$.  We consider the integral of $\e^{(1)}\o^{(1)}_0$ over $S^2$
and
$T^2$
separately.  The results must finally be assembled together with the correct
signs to find (\ref{3.50}).  (\ref{3.51}) will be considered after that.

Now $\e^{(1)}$ and $\o^{(1)}_{0}$ are closed on $\pa(\S\setminus \ch_1)$. The
integral of $\e^{(1)} \o^{(1)}_{0}$ over $S^2$ or $T^2$ is zero if either
$\e^{(1)}$ or  $\o^{(1)}$ is also exact.  A closed one form on $S^2$ being
exact, we thus have
\be
\int_{S^2} \e^{(1)} \o^{(1)}_0=0\,. \label{3.53}
\ee

As for $T^2$, let $\q^1,~\q^2$  be our canonical choice of coordinates thereon.
Then
\be
\e^{(1)}|_{T^{2}}= \a d \q^1 + \b d \q^2  + d \h^{(0)},~ \o^{(1)}_{0}|_{T^{2}}
=
\bar{\a} d \q^1 + \bar{\b} d \q^2+ d\bar{\h}^{(0)} \,,\;\;\; \a,~\b,~\bar \a,
{}~\bar \b = {\rm constants}. \label{3.54}
\ee
Here $d\h^0$ and $d\bar{\h}^{(0)}$ are exact.  Also one of the
coefficients $\bar{\a},~\bar{\b}$ for $\omega _0^{(1)}$ is zero for the case
in Fig. 9.
This is because one of the cycles on $T^2$ [the cycle not shown on $\pa \ch_1$
and $\pa \S$ in that figure] is homologous to a point, and a closed
one form on $\S \setminus \ch_1$ with a nonzero integral over this cycle does
not exist.  In any case,
\be
\int_{T^{2}} \e^{(1)} \o^{(1)}_{0}= (\a \bar{\b} - \bar{\a}\b)
\int_{T^{2}} d \q^1  d \q^2
= 4\p^2 (\a \bar{\b}-\bar{\a}\b)\, . \label{3.55}
\ee

We next look at (\ref{3.51}), again for $S^2$ and $T^2$ separately. $\e^{(0)}$
is a constant on $S^2$ or $T^2$.  Hence
\be
\int_{{S^{2}~{\rm or}~T^{2}}} \o^{(2)}_{0} \e^{(0)}= \e^{(0)}
\left|{}\right. _{{S^{2}~ {\rm or}~ T^{2}}}\times
\int_{{S^{2}~{\rm or}~T^{2}}} \o^{(2)}_{0}\, . \label {3.56}
\ee

We will now argue that the integral of $\o^{(2)}_{0} \o^{(1)}_{0}$ can be
chosen to be zero
in our examples.  Let us first consider Fig. 8. In that case, we can
choose $\o^{(2)}_{0}$ to be $d\q^1 d\q^2$ and $\o^{(1)}_{0}$ to be $d\q^1$ or
$d\q^2$.  Then $\o^{(2)}_{0} \o^{(1)}_{0}$ is zero and so is its
integral.

The result seems correct in the remaining case with $\S =\cb_3$ as well.  To
see this, we first
fill up $\S\setminus \ch_1$ with a family of tori and introduce coordinates
$\q^1,~\q^2$ for the tori and a coordinate $r$ labelling the tori.  Fig. 10
shows how this is done for $\S=\cb_3$. $r,~\q^1,\q^2$ give a coordinate
system for
$\S\setminus \ch_1$. $\o^{(2)}_{0}$ can then be identified with $d\q^1 d\q^2$
and $\o^{(1)}_{0}$ with $d\q^1$ or $d\q^2$ giving
$\o^{(2)}_{0} \o^{(1)}_{0}=0$. It may
be remarked that a torus touches in the middle of Fig. 10 so that a
cycle on
this ``torus'' is homologically trivial.  The corresponding form, say
$d\q^{2}$, is
thus ill defined on this torus. $\o^{(1)}_{0}$ is not this form, but $d\q^1$,
so that perhaps this singularity is not significant for our purposes.
\newpage

\noindent{\bf iii) Interpretation of $q_0(\a)$ and
$ P_N(\a)$}

We had previously used the integral of $B$ over a surface such as $S^2$
enclosing a
hole as a measure of the charge in the hole.  We had also used the integral of
$A$ on a loop $\cc$ as a measure of the flux through $\cc$.

But neither of these integrals is well defined either in the classical
canonical formalism or in quantum field theory.  This is because integrals of
fields over submanifolds are not meaningful in these contexts without
regularization or interpretation.

We now argue that $q_{0}(\a)$ is the correct expression for the charge
operator for charge in $\ch_\a$.  Similarly, $P_{N}(\a)$ is the corrected
version of
\be
\langle \int_{\cc_{N}} A \rangle \, . \label{3.57}
\ee

Let us first examine $q_{0}(\a)$.  According to (\ref{3.3}) and (\ref{3.11}),
\be
q_{0}(\a) = <\int_{\S\setminus \ch} d \x^{(0)}_{0} B >=
\frac {1}{\sqrt{\D}} <\int _{\pa \ch_{\a}} B - \int_{\S \setminus \ch}
\x^{(0)}_{0} d B >\, . \label {3.58}
\ee

\noindent The second term here is at least numerically zero in classical
physics, $dB$  being numerically zero in $\S\setminus \ch$ by (\ref{2.4}),
whereas the first term is the measure of charge in $\ch_\a$ we considered
in Section 2.1.  In this way, we can relate $q_{0}(\a)$ to the integral of
$B$. But this relation is necessarily imprecise since $q_0(\a)$ is well
defined whereas neither of the two terms in (\ref{3.58}) has a good meaning
in the canonical approach or quantum field theory.

In any case, when need arises, the first term in (\ref{3.58}) will hereafter be
identified with $q_{0}(\a)$.

As we saw earlier, $q_{0}(1)$ is canonically conjugate to $W(\o^{(2)})$:
\be
\{ <W(\o^{(2)})>,~q_{0}(1)\} =
\frac {1}{\sqrt{\D}} \int_{\pa \ch_{1}} \o^{(2)} \, . \label{3.59}
\ee

As for the interpretation of $P_{N}{(\a)}$, let us assume that $\S=\cb_3$ and
$\pa \ch_\a=T^2$.  The latter has coordinates $\q^i$ of which the $\q^2$ cycle
for fixed $\q^1$ can be taken to be $\cc_1$. [See Fig. 11.]
With $N=1$ and
$\a = 1$, we have
\be
P_1(1) =<-\int_{\S \setminus \ch} d \o^{(1)}_{1} A> =
<-\int_{\pa \ch_{1}} \o^{(1)}_{1} A -
\int_{\S \setminus \ch} \o^{(1)}_{1} d A>\, .  \label {3.60}
\ee
The last term is numerically zero by constraint.  Since
$\o^{(1)}_{1}|_{\pa \ch_{1}} = d \q^2$, we can write
\be
P_{1}(1) =<- \int_{\pa \ch_{1}} d \q^2 A_1 d \q^1 >,~
A|_{\pa \ch_{1}} : = A_1 d\q^1+A_2 d \q^2 \, . \label{3.61}
\ee

\noindent The integral
\be
\int A_1 d\q^1 \label {3.62}
\ee

\noindent over $\q^1$ is independent of $\q^2$, at least classically.  This is
because an
infinitesimal deformation of the $\q^1$ loop changes (\ref{3.62}) by an
integral of $dA$, and that is zero by constraint.  Thus $P_1(1)$ is
proportional
to the  flux through the $\q^1$ loop, or the flux on the vortex enclosed within
$\ch_1$.

\noindent{\bf 3.4. Summary of Observables and their Poisson Brackets}

We have encountered the following observables associated to $\pa \ch_\a$:
\be
{\rm Local ~Observables}~: q_n(\a),~p_n(\a),~P_N(\a) \, . \label{3.63}
\ee
\be
{\rm Nonlocal~ Observables}~: <W(\o^{(2)})>,~<V(\o^{(1)})> \, . \label{3.64}
\ee

\noindent Of these, $q_{0}{(\a)}$ is a measure of charge contained in
$\ch_{\a}$ and $P_{N}{(\a)}$ a measure of magnetic flux across the surface
with boundary $\cc_{N}{(\a)}$.  The nonlocal observables
$<W(\o^{(2)})>$ and $<V(\omega ^{(1)})>$ are conjugate to charge and
magnetic flux
respectively.  As we saw, they have an interpretation in terms of a Wilson line
integral involving $A$ and integral of $B$ over a surface.

It is enough to list the PB's involving these observables which are not
evidently zero.  They are as follows.
$$\{q_n(\a),~ p_m(\b)\}= \d_{\a\b} \d_{nm} \,,\;\;m\neq0\,,$$
$$\{q_n(\a),~P_N(\b)\} = 0\, ,$$
$$\{<W(\omega ^{(2)})>,~ q_n(1)\} = \sqrt{\D} \d_{n0}\, ,$$
$$\{  <V(\o^{(1)})>, ~ p_{\vec{N}}(\a)  \} = 0\,,$$
$$\{V(\o^{(1)}),P_N(\a ))\}=-\int_{\partial \ch _\a }
\o^{(1)}\o_N^{(1)}\, ,$$
\be
\{<W(\o^{(2)})>,~<V(\o^{(1)})>\} = \int_{\S\setminus\ch_{1}}
\o^{(2)} \o^{(1)} \, . \label{3.65}
\ee
An approach to the evaluation of the last integral here has also been outlined
in Section 3.3., ii).

\sxn{\bf Quantization, Diffeomorphisms and Vertex
Operators}{\label{sec-diffeo}

\noindent{\bf 4.1. Quantization and Diffeomorphisms}

For economy of notation, we will use the same notation for a quantum operator
and its classical counterpart.

The quantum version of (\ref{3.65}) is given by Dirac's prescription and reads

$$[q_n(\a),~ p_m(\b)]= i \d_{\a\b}\d_{nm} \, ,\sp m \neq 0 \, ,$$
$$[q_n(\a),~P_N(\b)] = 0\, ,$$
$$[<W(\o^{(2)})>,~ q_n(1)] = i\sqrt{\D} \d_{n0}\, ,$$
$$[<V(\o^{(1)})>, ~ p_{\vec{N}}(\a)] = 0\, ,$$
$$[<V(\o^{(1)})>, ~ P_N(\a)] = -\imath \int_{\partial \ch _\a}
\omega ^{(1)}\o _N^{(1)}\,,$$
\be
[ <W(w^{(2)})>,~<V(w^{(1)})>]=\imath \int _{\S \setminus \ch_1}
\o^{(2)}\o^{(1)}\,. \label{4.1}
\ee

As remarked previously, the operator algebra $\ca$ we realize on a Hilbert
space is generated by $q_{n}(\a),~ p_{n}(\a)$ and $P_N(\a).
<W(\o^{(2)})>$ and $<V(\o^{(1)})>$ will be treated in the way that the
Fubini-Veneziano field \cite{goddard} is treated in string theory.  Thus
suitably
regularized exponentials theoreof will be regarded as analogues of vertex
operators intertwining distinct representations of $\ca$.
Let us therefore temporarily set aside $<W(\o^{(2)})>$ and $<V(\o^{(1)})>$.

$q_{0}(\a)$ and $ P_{N}(\a)$ are in the center of $\ca$.  Their meaning in
terms of the charge and fluxes for $\ch_\a$ has been examined before.  These
operators are diagonal in a representation of $\ca$.  Thus if $|\cdot >$ is a
state in this representation space,
$$q_0(\a) |\cdot > = e_\a | \cdot > \,.$$
\be
P_N(\a) | \cdot > = F_N(\a) | \cdot > \, . \label{4.2}
\ee

\noindent These states are also of course annihilated by the Gauss law
constraints:
\be
\cg_j (\l^{(j)})|\cdot > = 0~ {\rm where }~ \l^{(j)}|_{\pa(\S\setminus \ch)}=0,
{}~~j = 1,2 \, .\label{4.3}
\ee

The algebra $\ca$ is the direct sum of  commuting subalgebras $\ca_\a,~\ca_\a$
having generators $q_n(\a),~p_n(\a)$ and $P_N(\a)$ for fixed $\a$:
\be
\ca = \oplus \ca_\a  \, . \label{4.4}
\ee

\noindent A representation of $\ca$ can thus be obtained by taking tensor
products of states carrying representations of $\ca_\a$.  A state in
(\ref{4.2}) and (\ref{4.3}) is such a product.  It is thus enough to consider
the representation of
one $\ca_\a$.  Let us  therefore fix the value of $\a$ for the present.

For $\ch_\a=\cb_3$ and ${\bf T}_3$, our bases for $\pa \ch_\a$ which define
the modes $q_n(\a)$, $p_n(\a)$ are shown in (\ref{3.14}) and (\ref{3.15}).
The basis elements enjoy the symmetries.
$$e^*_{Jm} = (-1)^m e_{J-m}\, ,$$
\be
e^*_{\vec{N}} = e_{-\vec{N}}\, . \label{4.5}
\ee

Let
\be
q_n(\a), p_n(\a) \rt q_{Jm}(\a),p_{Jm}(\a) \label{4.6}
\ee
when
\be
n \rt Jm \label {4.7}
\ee
and
\be q_n(\a),p_n(\a) \rt q_{\vec{N}}(\a),p_{\vec{N}}(\a) \label{4.8}
\ee
when
\be
n \rt \vec{N}\, . \label{4.9}
\ee

The reflection of (\ref{4.5}) are then the symmetries
$$q_{Jm}(\a)^*=(-1)^m q_{J,-m}(\a),~ p_{Jm}(\a)^*= (-1)^m p_{J,-m}(\a)\, ,$$
\be
q_{\vec{N}}(\a)^*=q_{-\vec{N}}(\a),~p_{\vec{N}}(\a)^*=p_{-\vec{N}}(\a)\, .
\label{4.10}
\ee

Let $\omega (\a):n\rt \omega _n(\a)(>0)$ be a frequency function invariant
under the substitution
$$n=Jm \rt n^*=J,-m$$
\noindent or
\be
n = \vec{N} \rt n^* = -\vec{N}\, . \label{4.11}
\ee

\noindent The dispersion relation is otherwise left arbitrary for the moment.

We can then construct the annihilation and creation operators
$$ a_n(\a)=
\frac{1}{\sqrt{2}} [\omega _n(\a)q_n(\a) + i p_n(\a)^\dagger ]\, ,$$
\be
a^{\dagger}_n(\a) =
\frac{1}{\sqrt{2}} [\omega _n(\a) q_{n}(\a)^\dagger - i p_n(\a)]\,.
\label{4.12}
\ee

\noindent Their only nonzero commutator is
\be
[a_n(\a),a_m(\a)^\dagger ]=\omega _n(\a) \d_{nm}\,. \label{4.13}
\ee

The algebra defined by (\ref{4.13}) can be realized on a Fock space in the
usual
way.

In our previous work, which examined the $BF$ system on $\S$ without holes,
the diffeomorphism (diffeo) group acting on $\pa \S$ was shown to be a group
of classical symmetries for the $BF$ system.  A generalized Sugawara
construction \cite{goddard} of its generators in terms of the analogues of
$q_n(\a),~p_n(\a)$ and
$P_N(\a)$ was derived.  It was argued that the full group of diffeos can not be
implemented on quantum states.  It was also shown that the group
$SDiff_0(\pa\S)$ of ``symplectic'' diffeormophisms can be implemented on
quantum states provided only that the frequency function $\omega (\pa \S):
n\rt \omega _n(\pa\S)$ appropriate to $\pa \S$ was independent of $n$ and
equal to a constant $\omega _0$.

Similar conclusions can be drawn with $\pa \ch_\a$ substituting for $\pa \S$.
Thus let $\h=\h^i\pa_i$ be a vector field which on $\pa \ch_\a$ is tangent to
$\pa \ch_\a$ and which vanishes on all other boundaries, and consider
\be
\ell (\h;\a) = \int_{\S\setminus \ch} (\cl _\h A) B \,.\label{4.14}
\ee

\noindent $\ell (\h;\a)$ generates the infinitesimal diffeo on $\pa \ch_\a$ for
the vector field $\h|_{\pa \ch_\a}$, the latter being the restriction of $\h$
to
$\pa \ch_\a$.   The PB's of $\ell (\h;\a)$ with $\cg_j(\l^{(j)})$ are weakly
zero so
that $\ell (\h;\a)$ leads to an observable.  It is also the case that
$\ell (\h;\a)\approx \ell (\h+\D\h;\a)$ if $\D\h|_{\pa(\S\setminus\ch )}=0.$
The equivalence class $<\ell (\h;\a)>$ of all $\ell $'s  weakly equal to
$\ell (\h;\a)$ can then be throught of as the observable generating the
infinitesimal diffeo of the vector field $\h|_{\pa \ch_{\a}}$ on $\pa\S$.

The rest of the analysis of $<\ell (\h;\a)>$ including its mode decomposition
and quantization follows \cite{we} with conclusions indicated above.

In \cite{we}, it was shown that the modes localized at $\pa \S$ can be
described by a
scalar field theory.  Following that work, it is easy to show that the modes
localized at $\pa \ch_\a$ can also be described by a scalar field theory.

\noindent{\bf 4.2. Vertex Operators }

\noindent{\bf i) Vertex Operator for Charge Creation}

Suppose that there is charge $e$ at $z_\a$ and we want to find an operator for
creating a state for this charge from one with
zero charge.  Such an operator is the vertex operator for charge $e$.

This vertex operator creates not just charge $e$ in $\ch_\a$, but also charge
$-e$ elsewhere.  This is true for the $BF$ system just as it is so for
Chern-Simons
dynamics \cite{bimonte2}.  Let us assume that the charge $-e$ is created at a
point $P$ on $\pa\S$.

We can proceed as follows to construct the state with these charges $e$ and
$-e$.  Enclosing $z_\a$, we have the associated hole $\ch_\a$, it being
understood that $\ch_\a$ will be shrunk to $z_\a$ at the end of the
construction.  Let us first assume that there is no charge in $\ch_\a$.  We
have thus just punched a hole at the point $z_\a$ where we want to create the
charge.

This hole $\ch_\a$, topologically a ball, has a boundary $\pa \ch_\a$ and an
associated Fock space of states $|\cdot ~~;z_\a>$ localized there and
carrying  zero charge:
\be
q_0(\a)| \cdot ~~;{z_{\a}}>= 0\,. \label{4.15}
\ee

\noindent Following earlier work on the Chern-Simons system \cite{moore}, we
can
interpret these states as describing spin fluctuations localized at $z_\a$
without a corresponding charge excitation.

The manifold $\S \setminus \ch$ in general has several connected boundaries
$\pa \ch_\b,~\pa\S$ and states localized thereon.  A physical state is a tensor
product constructed using one state from each connected boundary.  Let us
denote it by $| \cdot >$.
Let us also in particular denote any physical state which describes the Fock
vacuum at $z_\a$ by $|0>$.  In this notation, the dependence of $|0>$ on $z_\a$
has not been displayed.  In what follows, for specificity, we concentrate on
creating a charged state from $|0>$.

Next consider the Wilson line from the point $P$ on $\pa \S$ to $z_\a$, the
integral being along a line $L$:
\be
w(z_\a)=\exp ie \int_{P}^{z_{\a}} A \, . \label{4.16}
\ee

\noindent
Its response
\be
w(z_\a) \rt \exp [ie\x_{0}(z_\a)]w(z)\exp [-ie \x_{0} (P)] \label{4.17}
\ee
to the gauge transformation $A\rt A+d\x _0 $ shows that it creates a state
of charge $e$ at $z_\a$:
\be
{\sqrt{\D}}~q_{0}(\a) w(z_\a)|0> = e w(z_\a)|0> \, . \label{4.18}
\ee

\noindent [It also creates charge $-e$ at $P$.]  If the tangent to $L$ points
in the direction of the unit vector $\vec{n}$ from $z_\a$, then the spin of
this charge is localized for this state in direction $\vec{n}$.

Now for reasons mentioned earlier in Section 3.3 iii), the operator $w(z_\a)$
is not well defined, as it involves a quantum field on a line.  Using the
result (\ref{3.25}), we are thus led to consider
\be
\exp ie W(\bar{\o}^{(2)}) \label{4.19}
\ee
where $\bar{\o}^{(2)}$ has support on $L$.  This
$\bar \o^{(2)}$  is the same as the $\bar{\o}^{(2)}$ of Section 3.3 i) in the
limit $\d \rt 0$ and with the normalization constant $\tilde \l$ in
(\ref{3.26})
equal to 1.  As in Section 3.3, we will also assume for simplicity that there
is only one hole $\ch_\a$ for now, although a little later, we will have
occasion to briefly comment on the situation with several holes.

A few preliminary remarks are in order before studying (\ref{4.19}) further.
For a general topology, it is not correct to say $\o^{(2)}|_{\pa(\S\setminus
\ch)}$ determines $W(\o^{(2)})$ upto weak equivalence. [$\o^{(2)}$ need not
be
$\bar{\o}^{(2)}$.] For suppose that the closed two forms $\o^{(2)}$ and
$\o_0^{(2)}$ have both the same boundary values:
\be
\o^{(2)}|_{\pa(\S\setminus \ch)} =\o^{(2)}_{0}|_{\pa(\S \setminus \ch)} \, .
\label {4.20}
\ee

\noindent Then by the argument following (\ref{3.45}),

$$\o^{(2)}- \o^{(2)}_{0} = d\e^{(1)}\,,$$
$$d \e^{(1)}|_{\pa(\S\setminus \ch)} = 0\,,$$
or
\be
\e^{(1)}|_{\pa\S},~ \e^{(1)}|_{\pa \ch_\a} = ~{\rm closed~ one~ forms}~
\,.\label{4.21}
\ee

Suppose next that there is another one form $\bar{\e}^{(1)}$ with the same
boundary value as $\e^{(1)}$.
Then
\be
\int d (\e^{(1)}- \bar{\e}^{(1)})A = \int (\e^{(1)}-\bar{\e}^{(1)}) d A
\approx 0\, . \label {4.22}
\ee

\noindent Thus upto weak equivalence, $W(\o^{(2)} - \o^{(2)}_{0})$ is not
always zero, but is instead entirely
determined by $\e^{(1)}|_{\pa(\S \setminus \ch)}$.  Given this boundary value,
we are at liberty to chose its extension to $\S\setminus \ch$ in determining
the
equivalence class $<W(\o^{(2)}-\o^{(2)}_{0})>$.

Hence if there exists a closed one form in all of $\S\setminus \ch$ such that
its pull back to ${\pa(\S \setminus \ch)}$ is $ \e^{(1)}|_{\pa(\S \setminus
\ch)}$, then $W(\o^{(2)}-\o^{(2)}_{0})\approx 0$.  This is
for example the case if $\e^{(1)}|_{\pa(\S \setminus \ch)}$ is exact, so that
it can be written as  $d\h^{(0)}|_{\pa(\S\setminus \ch)}$.  We can then extend
$\h^{(0)}|_{\pa(\S \setminus \ch)}$ to a function $\h^{(0)}$ in
$\S \setminus \ch$ and
define the closed form extension of $\e^{(1)}|_{\pa(\S \setminus \ch)}$ to
$\S\setminus \ch$ to be $d\h^{(0)}$. This is also the case if for $\e^{(1)}$,
we can use a closed but not exact form of Section 3.3 ii).

These considerations show that $W(\o^{(2)}-\o^{(2)}_{0})$ is a linear
combination
of observables of type $P_N(\b)$ and similar observables $P_N(\pa \S)$ for
$\pa \S$ :
\be
W(\o^{(2)})-W(\o^{(2)}_{0}) = \sum_{N,\b} C_N(\b)P_N(\b)+\sum_{N}C_N(\pa\S)
P_N(\pa \S),~C_N(\b),~C_N(\pa \S)=~{\rm Constants}~\,. \label{4.23}
\ee

We can see (\ref{4.23}) in another way for those $\o^{(2)}$ with support on a
line $L$.  In that case, the statement that $<W(\o^{(2)})>$ is determined by
$\o^{(2)}|_{\pa(\S \setminus \ch)}$ is equivalent to the statement that it is
determined by the end points of $L$, and in addition by its tangents there when
they are at  the inner boundaries $\pa \ch_\a$. We can see that this statement
may be incorrect in the following way. In Fig. 12 (a,b),
we show
examples of $L$ with these same characteristics, but which differ by
noncontractible loops.  The corresponding $W$'s hence differ (upto constants)
by loop integrals of $A$, and the latter need not vanish.  Classically
they may
admit an interpretation in terms of fluxes
enclosed by the loops.

The uncertainties in  $W(\o^{(2)})$ as  represented by (\ref{4.23}) do not
affect its commutation relations.  The states created by (\ref{4.19}) differ
only by phases as a result of these uncertainties.  Therefore, to simplify
matters, let us assume for the present that $\S$ and $\ch$ are balls so that
this phase is absent.  Note that this $\ch $ is now also our $\ch_\a$.

Paranthetically, it may be remarked that these phases are important in
determining the statistical and other features of quantum states which depend
on the fundamental group of the configuration space.  In Fig. 12(b) for
example, the dotted loop can be interpreted as describing the transport of a
charge in a noncontractible loop around a vortex.  Section 5 will further
address this sort of issues.

We now return to (\ref{4.19}).  The specification of its action on $|0>$
requires the Fourier decomposition of $W(\bar{\o}^{(2)})$.  For now, we want to
concentrate on modes at $\pa \ch_{\a}$.  It is then best to change
$\bar{\o}^{(2)}$ to another closed two form $\O^{(2)}$, such that $W(\O^{(2)})$
commutes with all the observables localized
at $\pa\S$, except for the $\pa \S$
charge operator $q_0(\pa\S)$.[We will return to $\bar{\o}^{(2)}$ later.] The
method to achieve this objective is shown by (\ref{3.28}), (\ref{3.29}) and
(\ref{3.30}). Following those equations, we set
\be
\O^{(2)}\mid _{\partial \Sigma } = -\frac{1}{\D} \m \label{4.24}
\ee
[The minus sign arises from considerations involving orientation. For example,
the integral of $\bar \o^{(2)}$ over $\partial \ch_\a\bigcup \partial \S$ with
positive orientations must vanish by Stokes's theorem.]
For this choice, $W(\O^{(2)})$ describes a mode at $\pa\S$ which is conjugate
to
$q_0(\pa\S)$ and commuting with the remaining local observables at $\pa\S$.

We will continue to assume that the support of $\O^{(2)}$ in a neighbourhood
of $z_\a$ is on $L$.

With this choice of $\O^{(2)}$, we have deviated from an $\o^{(2)}$ with $L$
as support in all of $\S\setminus \ch$.  But earlier work \cite{bimonte2} shows
that
$\O^{(2)}$ is the one most appropriate for generalizing the Fubini-Veneziano
vertex operator \cite{goddard}.  We will return to the consideration of
$\bar{\o}^{(2)}$ later.

Since all $W(\o^{(2)})$ with the same boundary values for $\o^{(2)}$ are
weakly
equal for our chosen topology, the action of $e^{\imath eW(\o ^{(2)})}$
on a state is fixed
by $\o^{(2)}|_{\pa(\S\setminus \ch)}$. What remains thus for the specification
of
the action of $W(\O^{(2)})$ on a physical state is the display of its
dependence on the modes localized at $\pa \ch_{\a}$.  For this purpose, let us
first define the closed two form $\bar{\O}^{(2)}$ by
$$\bar \O^{(2)}|_{\pa \S} = -\frac{\m }{\D}\, , $$
\be
\bar \O^{(2)}|_{\pa\ch_{\a}}=\frac{\m }{\D}\,,  \label {4.25}
\ee

\noindent the two $\m's$ being the chosen volume forms on $\pa \S$ and $\pa
\ch_
\a$.
[They are not of course equal.  Also the details regarding $\bar{\O}^{(2)}$
away from the boundaries are not important here.]  Then
\be
(\O^{(2)}- \bar{\O}^{(2)})|_{\pa \S} = 0 \, , \label{4.26}
\ee
\be
(\O^{(2)}- \bar{\O}^{(2)})|_{\pa\ch_{\a}}  = d \x^{(1)}(\a)|_{\pa \ch_{\a}}\,,
\label{4.27}
\ee

\noindent (\ref{4.27}) following from the observation that the integral of its
left hand side over $\pa \ch_\a$ is zero.

After noting that there is no observable of type $P_N(\alpha )$ for
${\cal H}_\a = {\cal B}_3$,
we can expand $d \x^{(1)}(\a)|_{\pa \ch_{\a}}$ in a series of $e^*_N~\m$:
\be
d \x^{(1)}(\a)|_{\pa \ch_{\a}} = \sum_{N} a_N \;e^*_N ~\m \, . \label{4.28}
\ee

\noindent
The Fourier coefficients are given by
\be
a_M= \int_{\pa \ch_{\a}} e_M(\O^{(2)}-\bar{\O}^{(2)})\, . \label{4.29}
\ee

\noindent For $M=0$, this gives
\be
a_0 = 0\, . \label{4.30}
\ee

\noindent For $M\neq 0$, we find instead,
\be
a_M=\int_{S^{2}} e_M \O^{(2)} \, . \label{4.31}
\ee

Let us introduce the standard polar coordinates on $S^2$ and let $\O^{(2)}$
have support at $\q_0,~\f_0$. Then
\be
\O^{(2)}(\q,\f)|_{S^{2}}=\d(\cos\q-\cos\q_0)
\d(\f-\f_0)d \cos \q d \f \,. \label {4.32}
\ee
Using the correspondence $M \rt Jm,~e_M \rt \left(\frac{4\pi }{\D}
\right)^{1/2}Y_{Jm}$,
we have the following complete list of Fourier coefficients:
\be
a_{00}=0,~a_{Jm}=\left(\frac{4\pi }{\D}\right) ^{1/2} Y_{Jm}
(\q_0,\f_0)\,,\;\;\;{\rm for}~J\neq 0\,.
\label{4.32-b}
\ee

In this way, we find the mode decomposition
\be
<W(\O^{(2)}-\bar{\O}^{(2)})>=-\left(\frac{4\pi }{\D}\right)^{1/2}
\sum_{J,m;J\neq 0} Y_{Jm}(\q_0,\f_0)p_{Jm}(\a) \label{4.34}
\ee
or
\be
<W(\O^{(2)})>=<W(\bar{\O}^{(2)})> - \left(\frac{4\pi }{\D}\right)^{1/2}
\sum_{J,m;J\neq 0} Y_{Jm}(\q_0,\f_0)p_{Jm}(\a)\, . \label{4.35}
\ee

\noindent In this expansion, all but the first term are localized at
$\pa\ch_{\a
}$
while the first term is conjugate to charge at both $\pa \ch_\a$ and $\pa \S$.

The vertex operator for creation of charge $e$ at $z_\a$ can now be defined.
It is not quite (\ref{4.19}) with $\O^{(2)}$ for $\o^{(2)}$, but is its normal
ordered form as in string theory:
\be
\cw(\O^{(2)})=~:~e^{ie<W(\O^{(2)})>}~: \label{4.36}
\ee

\noindent The normal ordering is defined here by using the
creation-annihilation
 operators
of Section 4.1.

The Wilson line creates localized charge at both $\pa \ch_\a$ and
$\pa \S$.  It is thus associated to $W(\bar \o^{(2)})$, $\bar \o^{(2)}$
being supported on
$L$.  We must thus examine the mode expansion of
$W(\bar{\o}^{(2)}-\bar{\O}^{(2)})$. Since $\bar \o^{(2)}-\bar{\O}^{(2)}
|_{\pa \S} \neq 0$, the
expansion has an additional series of terms localized at $\pa \S$, similar
to the last group of terms in (\ref{4.35}).  Let $q_{Jm}(\pa\S)$,
$p_{JM}(\pa\S)$ be the modes localized at $\pa \S$ which are the counterparts
of $q_{Jm}(\a),~p_{Jm}(\a)$ and let $\Delta (\partial \S)$ be the area of
$\partial \S$ defined as in (\ref{3.1}). Then we find
$$<W(\bar \o^{(2)})>=<W(\bar{\O}^{(2)})> -
\left(\frac{4\pi }{\D}\right)^{1/2}\sum_{J,m;J\neq 0}Y_{Jm}(\q_0,\f_0)
p_{Jm}(\a)$$
\be
+ \left(\frac{4\pi }{\D}\right)^{1/2}\sum_{J,m;J\neq 0}Y_{Jm}(\q_0^{\prime},
\f_0^{\prime})p_{Jm}(\pa \S)\,.\label{4.37}
\ee

\noindent Here, we have introduced polar coordinates $\q^\prime,~\f^\prime$ on
$\pa \S$ and $\q_0^\prime,~\f^\prime_0$ is the point where $L$ joins $\pa\S$.

We thus have the result
\be
{\rm Regularized~ Wilson~ line}~=: e^{ieW(\bar \o^{(2)})}:
\label {4.38}
\ee

The construction leading to (\ref{4.36}) and (\ref{4.38}) has relied on
particular choices of $\S$ and $\ch$.  For a more general situation, we can
proceed as follows.  Given a direction at $z_\a$ and a point $P$ on
$\partial \S $, we first
choose a particular line $L$ from $P$ to $z_\a$ with its tangent at $z_\a$
being in the given direction.  We then define the mode analysis following what
we did above, assuming that the $P_N(\a)$ terms are absent, say. [The basis
functions $e_m^{(\a)}$ should of course be appropriate for the topologies of
$\pa \S$ and $\pa \ch_\a$.  Also there is no way to tell whether or not there
are such terms for a given $L$, their choice being a convention.]  The
regularized Wilson integral or the vertex operator can then be constructed.  If
there is another line $L^\prime$ and $L^\prime$ also originates at $P$ and ends
at $z_\a$ with the same direction of tangent, then $L$ can be smoothly changed
to $L^\prime$ keeping its end at $P$ fixed, but changing its other end
smoothly.  The vertex operator for $L^\prime$ can then be expressed as the one
for $L$ plus factors involving $P_N(\b)$. [Cf. (\ref{4.23}).]  This adiabatic
transport also relates the corresponding states they create from $|0>$.  Note
that although the $P_N(\a)$ factors for one $L$ is a matter of choice, the
additional such factors created when $L$ is
changed $L^\prime$ can be determined and has an intrinsic meaning.

The classical configuration of charges and vortices is specified by their
locations, charges, fluxes and spin directions.  $L$ and $L^\prime$ are thus
both associated with the same point of this configuration space $Q$, and they
give states which differ by a phase.  The smooth deformation of $L$ to
$L^\prime$ corresponds to parallel transport of quantum states in a loop in
$Q$.  The phase above is thus the holonomy for this loop, and the quantum
states are really to be thought of as sections of vector bundles over $Q$.

\noindent{\bf ii) Vertex Operator for Vortex Creation}

Let us suppose that there is a magnetic vortex along a loop $\cc_\a$ which we
assume for simplicity to be an unknot.  Our task is to find an operator which
creates this loop just as (\ref{4.36}) creates charge.

We can follow Section 4.2.~i) in order to define this operator.  Thus we begin
with a manifold  where no magnetic vortex is present at $\cc_\a$ and then
punch a hole $\ch_\a$
enclosing $\cc_\a.~\ch_\a$ is a solid torus and it is eventually to be shrunk
to
$\cc_\a$.

The hole $\ch_\a$ has boundary $\pa \ch_\a$ and Fock states
$|\cdot ~~;{\cc_\a}>$
localized there.  They are states which carry zero magnetic flux.  Just as for
the charge problem, they describe spin fluctuations which are not associated
with a flux excitation.  Let $|0;{\cc_\a}>$ be the vacuum state at
$\pa \ch_\a$.

The manifold $\S\setminus \ch$ in general has many boundaries and there is a
family of Fock states localized on each of its connected components.  A
physical state is a tensor product formed of these states as described in
Section 4.2.i).  As before, let us call a physical state, with the vacuum at
$\pa \ch_\a$ as a factor, as $|0>$, suppressing its dependence on  $\cc_\a$.

Let $S_\a$ be a surface with $\cc_\a$ as boundary as in Fig. 13 and consider
\be
e^{-i{\F_{\a}} \int_{S_ \a}B} | 0 > \, . \label{4.39}
\ee

The operator which measures flux on $\cc_\a$ is the integral
\be
\int_{L_{\a}}{A} \label{4.40}
\ee

\noindent along the loop $L_\a$ of Fig. 13.  As $A$ and $B$ are conjugate
operators, we find that the state (\ref{4.39}) describes a vortex of flux
$\F_\a$:
\be
\left(\int_{L_{\a}} A\right) \left(e^{-i{\F_{\a}} \int_{S_{\a}}{B}}
\right)|0>= \F_\a
e^{-i\F_{\a} \int_{S_{\a}}{B}}|0>  \, . \label{4.40-b}
\ee

\indent Just as (\ref{4.16}), the operator in (\ref{4.39}) is not well
defined. We are thus led to consider the exponential constructed from
\be
V(\bar{\o}^{(1)}) = \int \bar{\o}^{(1)} B \, , \label{4.41}
\ee

\noindent $\bar{\o}^{(1)}$ being a closed one form with support on $S_\a$
and
\be
\int_{L_{\a}} \bar{\o}^{(1)} = 1 \, . \label{4.42}
\ee

Our next task is to examine the dependence of (\ref{4.41}) on the choice of
$S_\a$ for a given $\cc_\a$.  For continuous deformations of $S_\a$ with
tangent directions to $S_\a$ at $\cc_\a$ held fixed, $V(\bar{\o}^{(1)})$
changes
only by constraints as an application of Stokes' theorem shows.  But there are
in general surfaces $S_\a$ and $S^\prime_\a$ which are not mutually homotopic
and which have same boundaries and
tangent directions there.  An example is shown in Fig. 14.
The two surfaces in this figure give $V(\bar{\o}^{(1)})$ differing by a term
proportional to the charge operator $q_0(\b)$ and this term need not be zero
if the $\pa \ch_\b$ state has nonzero charge.

We can understand such ambiguities in general by considering $V(\o^{(1)})$ and
$V(\o^{(1)}_{0})$ where the closed forms here need not be supported on
surfaces.
The only condition we will impose is that their pull backs on boundaries agree:
\be
\o^{(1)}|_{\pa(\S\setminus \ch)} = \o^{(1)}_{0}|_{\pa(\S\setminus \ch)}\, .
\label{4.43}
\ee

\noindent Our task is to determine the nature of $\o^{(1)}-\o^{(1)}_{0}$ given
(\ref{4.43}).

Using what follows (\ref{3.45}), we can conclude from (\ref{4.43}) that
$$\o^{(1)}-\o^{(1)}_{0} = d \e^{(0)}\,,$$
$$d \e^{(0)}|_{\pa(\S \setminus \ch)} = 0\,,$$
or
\be
 \e^{(0)}|_{\pa\S} ,~ \e^{(0)}|_{\pa \ch_{\a}}= ~{\rm constant ~functions}~
\, . \label{4.44}
\ee

Now if $\e^{(0)}$ and $\bar{\e}^{(0)}$ have the same boundary value,
then
\be
\int d (\e^{(0)}- \bar{\e}^{(0)}) B \approx 0 \, . \label{4.45}
\ee
Hence, upto weak equivalence, $V(\o^{(1)} - \o^{(1)}_0)$ is determined
by $\e^{(0)}|_{\pa (\S \setminus \ch)}$.  Given this boundary value, we are
at liberty to choose any one of its extensions in determining the eqivalence
class $<V(\o^{(1)}-\o^{(1)}_{0})>$.

Since $\e^{(0)}$ is a constant on each connected component of the boundary, it
follows that
$$<V(\o^{(1)} - \o^{(1)}_{0})> =
\sum_{\b} D(\b) q_0(\b) +
D(\pa \S) q_0(\pa \S),$$
$$D(\b) =\sqrt \D\e^{(0)}|_{\pa \ch_{\b}} \, ,$$
\be
D(\pa \S) =\sqrt \D\e^{(0)}|_{\pa \S} \, . \label{4.46}
\ee

\noindent (\ref{4.46}) is similar to (\ref{4.23}).  Just as in that case, such
extra terms $<V(\o^{(1)}-\o_0^{(1)})>$ in $V$ do not affect its commutation
relations with local
observables, $q_0(\b)$, $q_0(\partial \S )$ being superselected operators.
Nevertheless they are
important in determining adiabatic transport properties of states, just as in
Section 4.2 i).

Our next task is the mode decomposition of $V(\bar{\o}^{(1)})$.  For this
purpose, given a $\cc_\a$ and a field of directions at $\cc_\a$, we first
choose
an $S_\a$ with $\pa S_\a= \cc_\a$ and with its tangents at $\cc_\a$ pointing in
the given directions.  For this $S_\a$, we then arbitrarily assume that we can
ignore charge terms like those in (\ref{4.46}), this assumption amounting to
the choice of a phase
convention just as in the analogous situation in Section 4.2 i).  The Fourier
analysis of $<V(\bar{\o}^{(1)})>$ can then be accomplished by Fourier analysing
$\bar{\o}^{(1)}|_{\pa \ch_\a}$ using the differentials $d e_{\vec{N}}$ of our
chosen basis.

Let us choose coordinates $\q^i$ on $\pa \ch_\a= T^2$ as indicated in Fig. 11.
Then $\bar{\o}^{(1)}|_{\pa \ch_{\a}}$ is supported on a loop
$S_\a \bigcap \pa \ch_\a$ with coordinates $(\q^1(\q^2),\q^2)$. [ Here
we are assuming for clarity
that $\ch_\a$ has a finite cross section, although finally we must let it
become a point.]  For reasons we will see later, we can not think of a neat
way of
explicitly displaying the mode analysis for a general dependence of $\q^1$ on
$\q^2$.  Let us therefore assume that $\q^1$ has the same value $\q^1_0$ for
all
$\q^2$:
\be
\q^1(\q^2)= \q^1_0 \, . \label{4.47}
\ee

Let $\O^{(1)}$ be a closed one form which has the following
properties:
$$\O^{(1)}|_{\pa \ch_{\a}} = \frac {d\q^{1}}{2\p} \, ,$$
\be
\O^{(1)}|_{\S\setminus \bar{\ch}_{\a}}= \bar{\o}^{(1)}|_{\S\setminus
\bar{\ch}_{\a}} \, . \label{4.48}
\ee
Here $\bar \ch_\a$ is a solid torus enclosing $\ch_\a$ (and no other
holes) as in
Fig. 15.
Such an $\O^{(1)}$ is readily seen to exist.  Now since
\be
(\bar{\o }^{(1)}-\O^{(1)})|_{\pa \ch_{\b}~{\rm or}~\pa \S}=0,~~\b\neq \a \,,
\label{4.49}
\ee
$V(\bar{\o}^{(1)}-\O^{(1)})$ depends upto weak equivalence [and charge
terms which will be set equal to zero] only on
$(\bar{\o }^{(1)}-\O^{(1)})|_{\pa \ch_{\a}}$.

The mode expansion of $(\bar{\o}^{(1)}-\O^{(1)})|_{\pa \ch_{\a}}$ can be
accomplished by first writing
\be
(\bar{\o}^{(1)}-\O^{(1)})|_{\pa \ch_{\a}} = d \x^{(0)}(\a) \, . \label{4.50}
\ee
We then expand $\x^{(0)}(\a)$ assuming that its constant mode is absent:
\be
\x^{(0)}(\a) = \sum_{\vec{N}\neq 0} b_{\vec{N}} e_{\vec{N}}\, . \label {4.51}
\ee

\noindent Hence
\be
d\x^{(0)}(\a)=\imath \sum_{\vec{N}\neq 0} b_{\vec{N}}e_{\vec{N}} N_j d \q^j
\, . \label{4.52}
\ee

\noindent This is also equal to
\be
(\bar{\o}^{(1)}-\O^{(1)})|_{\pa \ch_{\a}} = \d(\q^1-\q^1_0)d\q^1-\frac
{d\q^{1}}{2 \p} \, . \label{4.53}
\ee
Therefore
$$b_{\vec{N}} = 0~ {\rm for}~ N_2 \neq 0 \, ,$$
\be
b_{N_{1},0} = -\imath \frac{\sqrt \D}{2\pi N_{1}} e^{-iN_{1}\q^{1}_{0}}~{\rm
for}~ N_1
\neq 0 \, .\label{4.54}
\ee

We thus find,
\be
<V(\bar{\o}^1)>=<V(\O^{(1)})>-i\frac {\sqrt \D}{2\p }\sum_{N_{1}\neq 0}
\frac {e^{-iN_{1}\q^{1}_{0}}}{N_1} q_{N_{1},0}(\a )  \,. \label{4.55}
\ee

\noindent In this expansion, all but the first term are localized at
$\pa\ch_{\a}$, whereas the first term can change  by charge terms if the
surface $S_\a$ entering its definition is changed to another homotopically
different surface.

The vertex operator for the creation of a vortex is
\be
{\cal V}(\bar \o^{(1)}):=\,\, :e^{-\imath {\F _\a} <V(\bar \o ^{(1)}>}:\,,
\label{4.57}
\ee
the creation and annihilation operators of Section 4.1 being used to
define the normal ordering.

If the vortex were more general than what (\ref{4.47}) describes and has
coordinates
$(\q^1(\q^2),\q^2)$, the clumsiness in analysis  will occur when we try to
compute $b_{\vec{N}},~\q^1_0$ in (\ref{4.53}) having to be replaced by
$\q^1(\q^2)$. Although integral formulae for $b_{\vec{N}}$ can be readily
written down, they can not always be neatly evaluated.

Remarks similar to those in Section 4.2 i) leading upto the assertion that
states are sections of vector bundles apply with equal force here as well.

\sxn{\bf Spin and Statistics, Aharonov-Bohm Interactions}
\label{sec-statistics}

\noindent {\bf 5.1. Preliminares}

If $\s$ is the operator for the exchange of two identical constituents of a
system, and $R_{2\p}$ is the operator for the $2\p$ rotation of one of the
constituents, the spin-statistics theorem asserts that these two are identical
operators on  quantum states:

\be
\s =R_{2\p} \, . \label {5.1}
\ee

For extended systems such as a vortex \cite{bal}, or for dynamics supported on
an underlying manifold with nontrivial connectivity \cite{imbo}, there are in
general
several distinct ways of performing the exchange.  The exchange operator $\s$
in (\ref{5.1}) then corresponds to the adiabatic transport of the constituents
when they are confined in a contractible open set (say the interior of a ball
$\bar \cb_3 $) of the underlying manifold. The open set is assumed to contain
only these constituents, and
certain nontrivial motions available for extended systems are also excluded.

In Section 5.2, we will discuss how (\ref{5.1}) can be proved for charges and
vortices in the $BF$ system, explaining at the same time the particular
exchange
transport which enters (\ref{5.1}).  In Section 5.3, we then establish a
genuinely new spin-statistics theorem for vortices which shows the identity of
two operators on quantum states.  The first is the operator $\hat{\s}$ of
interchange \cite{aneziris,ajit2}, the latter involving an exhange transport
of vortices which
is not the same as the one for $\s$.  It is explained by figures in Section
5.3.  As for the second, let  the loop $\cc$ be the location of the vortex in
$\S$.  For a state created by a vertex operator such as (\ref{4.57}), we can
then associate a (spin) direction to each point $p$ of $\cc$. Let
$\Hat{R}_{2\p}$ denote the $2\p$
 rotation of all these directions around their $p$'s,  the axis
of rotation being the tangent to $\cc$ at $p$. [It is also illustrated in
Section 5.3.] Let us call this operator as ``internal $2 \p$ rotation.''   The
result we find is then the identity of $\hat{\s}$ and $\Hat{R}_{2 \p}$:
\be
\hat{\s} = \Hat{R}_{2 \p} \, . \label{5.2}
\ee

There is one remarkable feature of these theorems which merits emphasis here.
The possibility of creation-annihilation processes was an important ingredient
in certain earlier work on spin and statistics including our own
\cite{bal3,balclinn}.  In contrast, the proofs
constructed here do not seem to use creation-annihilation processes, at least
in any manifest way.  Further understanding of these apparently distinct proofs
is thus indicated.

In Section 5.4 we briefly consider
what may be called the Aharonov-Bohm interaction of a charge and a vortex.
The interaction phase comes about when the charge is transported in a loop
enclusing the vortex flux, much as in the usual Aharonov-Bohm effect. This
short
discussion is included here because the spin-statistics theorems too are
associated with transports of charges and vortices.

The discussion in this Section will assume for simplicity that the charges
and vortices have sharp spin states, that is that they are created at
$\partial \ch_\a$ by vertex operators like (\ref{4.36}) and (\ref{4.57}).
\newpage

\noindent {\bf 5.2. The Standard Spin-Statistics Theorem}

\noindent {\bf i) Identical Charges}

Let us suppose that the charges are located at positions $z_1$ and $z_2$ in
$\S$ and enclosed by small balls $\ch_1$ and $\ch_1$, their radii eventually
becoming zero.  There may also be other charges and vortices in $\S$ in
addition to these charges.

Without loss of generality, we can assume that $z_1$ and $z_2$ are located in
the interior of a ball $\bar{\cb}_3$ which excludes all other charges and
vortices, and does not touch $\pa\S$, as shown in Fig. 16.
The state of these two charges with identical internal states can then be
written as
\be
(:\exp i e \int_{L_{1}} A :)(:\exp i e \int_{L_{2}} A:) |0> \label{5.3}
\ee

\noindent
the tangents to $L_i$ at $z_i$ being parallel. [This parallelism of vectors
and hence identity of internal states at distinct points can be defined as
follows.  We first fix a flat metric in the interior of $\S$.  We then use its
connection to parallel transport vectors.  If directions of two vectors are
related by this parallel transport, we declare them to be parallel.]

The Wilson integrals of (\ref{5.3}) are the regularized Wilson integrals of
Section 4.2 i). [Cf. (\ref{4.38}).]   The state (\ref{5.3}) also contains
``image'' charges at $z_i^\prime$ (shown in Fig. 16)
which for convenience we assume are located
in the interior of $\S$.  The positions and internal states of these image
charges will be held fixed throughout the considerations below.

The spin-statistics theorem can now be proved following \cite{bimonte}.  If
(\ref{5.3}) is
represented by Fig. 16, then the theorem here is the identity of Fig. 17. [In
Fig. 17, $\S$ and holes except those of $z_i$ and $z^\prime_i$ are omitted.]

For completeness, we next show that the $2\p$ rotation of the charge at $z_2$
is actually trivial.  In other words, $R_{2\p}$ and hence $\s$ are unit
operators, and the charges are integral spin (or tensorial) bosons.  This
result is not surprising since according to the remarks in  Section 4.1, these
charges are described by scalar fields.

The proof that $R_{2\p}=1$ is accomplished by continuosly changing
$L^\prime_2$ to the configuration $L^{\rm''}_{2}$ of Fig. 18
without ever
changing its end points or tangents there, or touching $\ch_2$.  This can be
done by first lifting and then sliding the interior of $L_2$ Fig. 18, the
dotted portion there lying well above $\ch_2$ and then shifting the lifted
portion so as to get $L^{\rm''}_2$
\newpage
\noindent
{\bf ii) Identical Vortices}

The vortices are assumed to occupy loops $\cc_1$ and $\cc_2$ and to be enclosed
in a ball $\bar{\cb}_3$ which contains no other charge or vortex and which does
not touch $\pa\S$. They are to have identical internal ``spin states'' created
by regularized vertex operators of type (\ref{4.57}).  The identity of vortices
is defined here as follows. We choose a flat metric and its connection within
$\S$ as in Section 5.2 i) above.  Now there are several ways we can
continuously
being illustrated in Fig. 19.

Suppose we can find one such special motion with
the following property: A point $p_2$ of $\cc_2$ for any of these motions
traces a curve $L$ and ends up at a point $p_1$ of $\cc_1$.  Consider the
parallel transport of the internal vector at $p_2$ along $L$ to $p_1$.  Then
for this special motion, this vector at $p_1$ must be parallel to the internal
vector of the $\cc_1$ vortex at $p_1$.  If one such special motion can be
found, we will say that these two vortices have identical internal states.

It is convenient to display the two disks (assumed circular for the chosen
metric) by taking their sections along the equator.  Fig. 19(b) represents
Fig. 19(a) in this fashion.

In Fig. 19(a), the shaded regions are the surfaces over which $B$ is
integrated while constructing the vertex operators.

In Fig. 20, we have displayed a path for the adiabatic exchange of two
identical loops.  The final state is exactly the same as the initial state.
Therefore, defining $\s$ to be the operator producing this final state from the
initial state, we find
\be
\s = 1 \, . \label{5.4}
\ee
These loops are thus bosons.

Figure 21 shows the path for $2\p$ rotation of a vortex.  $R_{2\p}$ is the
operator producing the final state in this figure from the initial state.  But
clearly the final and initial states are identical, and therefore
\be
R_{2\p} = 1 \label{5.5}
\ee
Thus the vortex has integral spin or tensorial states.  Also the
spin-statistics theorem (\ref{5.1}) is satisfied.

\newpage
\noindent{\bf 5.3. A New Spin-Statistics Theorem: Interchange=Internal $2\p$
Rotation}

Interchange \cite{ajit2} is the operation of exchanging identical vortices
where one
vortex is first taken through the middle of the other vortex.  Representing a
vortex by a loop, this motion is illustrated in Fig. 22.
$\hat \s$ is the
operator producing the final state here from the initial state.

The internal $2\p$ rotation describes $2\p$ rotation of the spin directions at
every point $p$ of the vortex around the tangent to vortex at $p$.  The history
of a spin direction at a particular point $p$ is illustrated in Fig. 23.

Figure 24 shows the interchange on a two-vortex state in detail including the
way we can distort the surfaces as this ``adiabatic'' process is being
performed.  The interior of the surfaces should not touch
vortex locations, and for this reason, the
surfaces must necessarily be deformed during the process. In the passage from
(d) to (e), we have distorted a surface in the direction of the double
arrows till
it touched itself, and then pinched off the resultant bubble.  In going from
(g) to (h), we have deformed a surface in the direction of the double arrow
till its middle portion touched the middle portion of the other surface.  The
integral of $B$ over these middle portions cancel leading to (h).

Figure 24(h) shows that interchange is internal $2\pi$ rotation of spin frames
establishing (\ref{5.2}).

There is an operation called slide defined in \cite{aneziris} and illustrated
in
Fig. 25 using a presentation similar to Fig. 22.  In this operation, the left
vortex is spatially stationary whereas the right vortex is taken in a loop
which passes through the middle of the left one.
The operator producing
the final state of this sequence from the initial state is the operator $S$ of
slide.  It is easy to see that the path which is the composition of slide and
exchange is interchange. Hence
\be
\hat{\s} = \s S \,. \label{5.6}
\ee
But as $\s=1$ by (\ref{5.4}), the result (\ref{5.2}) implies that
\be
S= \Hat{R}_{2\p} \,. \label{5.7}
\ee

Internal spin directions had occurred in an earlier work on spin-statistics
theorem for vortices \cite{balclinn}.  It is significant that Srivastava
\cite{ajit3} has
succeeded in proving (\ref{5.2}) using the approach of that paper and without
appealing to the $BF$ Lagrangian used in this paper.
\newpage
\vspace{10mm}
\noindent{\bf 5.4. Aharonov-Bohm Interactions }

Let $|0>$ be the "vacuum" state as defined previously. Let us apply vertex
operators to $|0>$ to create therefrom a vortex at ${\cal C}_\alpha $ and a
charge at $z_\beta $ as in Fig. 26(a). The figure also shows the surface
$S_\alpha $ and the line $L_\beta $ involved in the definition of vertex
operators, the state being
\be
|\Psi _I>=:e^{\imath e_\beta \int _{L_\beta }A}::e^{-\imath \Phi _\alpha \int
_{S_\alpha }B}|0> \label{5.8}
\ee
(The figure does not show other charges and vortices, the holes or $\partial
\Sigma $.)

Now consider the transport of the charge in a loop $\tilde L$ around ${\cal C}
_\alpha $ without changing its spin state. This loop is shown in Fig.26(b),
The sucessive stages of (\ref{5.8}) for this transport can also be realized as
in Figs.26(c-h).The passage from Fig. 26(d) to Fig. 26(h) should be clear. Now
the integral of $A$ over a line does not change if it is distorted in its
interior without crossing
${\cal C}_\alpha $. By the equality of Fig. 26(d) and 26 (e), this means that
$L'$ can pierce the surface in Fig. 26(e) anywhere without changing the state.
Note that for Fig. 26(h), the $B$ integral goes over $S_\alpha $ as well as
over
the bubble with surface $S_\alpha ''$ enclosing the charge. When the transport
around $\tilde L$ is completed, we do not recover $|\Psi _I>$, because of the
additional phase from the integral of $A$ over $L''$. The integral of
$B$ over
$S_\alpha ''$ gives no additional phase as the $B$ integral in (\ref{5.8}) is
next to $|0>$ and the $dB$ integral over the ball enclosed by $S_\alpha ''$
annihilates $|0>$. The additional phase in question follows from
(\ref{4.40-b}) so that
\be
|\Psi _I> \stackrel{{\rm on~transport~around}~\tilde L}
{\longrightarrow}e^{\imath e_\b\Phi_\a}|\Psi _I>\label{5.9}
\ee

The state $|\Psi _I>$ can equally well be written with the factors involving
$A$ and $B$ interchanged:
\be
|\Psi _I>=:e^{-\imath \Phi _\alpha \int _{S_\alpha }B}:\:
	  :e^{\imath e_\beta \int _{L_\beta }A}:\,|0>. \label{5.10}
\ee

For this form of $|\Psi _I>$, it is the phase from the integral of $A$ over
$\tilde L$
which becomes 1 whereas the integral of $B$ over $S_\alpha ''$ leads to
(\ref{5.9}).

The phase in (\ref{5.9}) becomes 1 and the Aharonov-Bohm interaction vanishes
if the quantization condition
\be
e_\beta \Phi _\alpha = 2\pi\times {\rm Integer} \label{5.11}
\ee
is fulfilled.

Suppose now that we replace the vortex in (\ref{5.8}) with a charged vortex
with
charge $e_\alpha $ and flux $\Phi _\alpha $. Suppose also that the charge in
(\ref{5.9}) is also replaced by a charged vortex with charge $e_\beta $ and
flux $\Phi _\beta $. The state $|\Psi _I>$ then becomes
\be
|\Psi _I'>=:e^{-\imath \Phi _\beta \int _{S_\beta }B}:
:e^{\imath e_\beta \int _{L_\beta }A}:
:e^{-\imath \Phi _\alpha \int _{S_\alpha }B}::e^{\imath e_\alpha
\int _{L_\alpha }A}:|0> \label{5.12}
\ee
where $S_\alpha $, $L_\alpha $, $S_\beta $, $L_\beta$ are shown in Fig.27. The
transport around $\tilde L$ now becomes a slide and the phase change is readily
computed to be $e^{\imath e_\beta \Phi _\alpha }$,
\be
|\Psi _I'>\stackrel{\rm After~a~slide}{\longrightarrow}e^
{\imath e_\beta \Phi _\alpha }|\Psi _I'>.\label{5.13}
\ee

It thus becomes 1 if
\be
e_\beta \Phi _\alpha =2\pi\times {\rm Integer}\, .\label{5.14}
\ee
(\ref{5.14}) is the analogue of Dirac quantization condition for dyons
\cite{dyons}.

\sxn{\bf Self Energies of Sources and When They Diverge}\label{sec-energies}

In our earlier work \cite{we}, we have discussed the edge states of the
Lagrangian
$L_0$ of (\ref{1.6}). This Lagrangian has the virtue of including the
Maxwell Lagrangian of $A$ and the corresponding Lagrangian of $B$.  In that
paper, it was shown that the structure of edge states was not sensitive to
these terms and that they occur equally well in $L_0$ and $L_0^*$.

The edge states treated in this paper are those at $\pa \S$.  There are also
these states at source boundaries, and as mentioned in the Introduction, they
too are present in the Lagrangian
\be
L=L_0+L_I \label{6.1}
\ee

\noindent obtained by replacing $L^*_0$ in (\ref{1.13}) by $L_0$.

But as stated in the Introduction, the Hamiltonian for (\ref{6.1}) diverges
in the presence of sources much as in electrodynamics.  Here we briefly
indicate how this happens for (\ref{6.1}) for charges.  Divergences are present
for vortices too much as in the work of Lund and Regge \cite{lund}.

The Hamiltonian for (\ref{6.1}) was derived in \cite{we} and reads

\begin{eqnarray}
H & = &\int d^3x\left[ \frac 12\left[\pi _i -\epsilon _{ijk}B_{jk}\right]^2+
     \frac {\lambda }{16}P^2_{ij}
     + \frac 14F_{ij}^2 + \frac {1}{3\lambda }H_{ijk}^2 \right. \nonumber \\
  &   &  - A_0(\partial _i\pi _i-e\delta ^3(x-z)) -B_{0i}
(\partial _jP_{ji}+ 2\epsilon _{ijk}\partial _jA_k - \lambda
\int d\sigma ^1\frac{\partial y^i}{\partial \s^1}\delta ^3(x-y)\nonumber \\
  &   &\left. +\psi ^0\pi _0+\psi ^iP_{0i}\right]\,,\label{6.2}
\end{eqnarray}
the coefficients of $A_0$ and $B_{0i}$ being constraints.

Now in the presence of charge, $\partial _i \pi _i$ has a $\d$-function
singularity and hence $H$ is classically divergent.

There is a similar divergence in quantum theory too.  Thus suppose that $H$ has
been properly normal ordered and vanishes on the state $|0>$ of Section 4.2 i),
$|0>$ having no charge or vortex. It is thus annihilated in particular by the
operator $\partial_i\pi_i$ associated with the
Gauss law constraint for $e=0$. Then a state with a single
charge, such as
\be
W(\Omega ^{(2)})|0> \label{6.3}
\ee
is annihilated only by the operator associated with the Gauss law constraint
containing also the point charge contribution. $H$ is therefore divergent on
this state.

\sxn{\bf Twisted Spins on Vortices}

There is a spin direction attached at each point of the vortex and the latter
topologically is a circle $S^1$. We can thus conceive of the spin direction
rotating
by $2\pi $ as one goes around the vortex as illustrated in Fig. 28.
More
generally, we can conceive of this direction rotating by $2\pi N$
($N\in {\bf Z}$) as one goes around the vortex, thereby suggesting the
configuration of a soliton for winding number $N$. In this concluding Section,
we briefly discuss when vertex operators can be found for such twisted spins.

First we explain the precise definition of the winding number of the vortex
spin. We assume for simplicity that the vortices have sharp transverse spins.
Now, if we enclose a vortex inside a solid torus ${\bf T}_3$ in the usual way,
then there are two cycles $Y_1$ and $Y_2$ on its boundary
$T^2=\partial {\bf T}_3$, $Y_i$ corresponding to $\q_i$ of Fig.2 increasing
by $2\pi $. If lines (geodesics) are drawn in the direction of the transverse
spins, they will pierce $T^2$ along the curve $Y$ which is homologous to the
curve obtained by traversing $Y_1$ $N$ times and $Y_2$ once, $Y\sim NY_1+Y_2$.
The integer $N$ is then defined to be the winding number of the vortex spin.
[Homologous curves are here defined by regarding them as curvs on $T^2$ and
not in ${\bf T}_3$.]

Referring to Fig. 13 or Eq. (\ref{4.39}), we see that the construction of a
vertex operator for
a vortex with spatial location ${\cal C}_\alpha $ involves the existence of a
surface $S_\alpha $ with boundary $\partial S_\alpha ={\cal C}_\alpha $ and
with directions of its tangents at ${\cal C}_\alpha $ giving the spin
directions. Such a surfaces $S_\alpha $ must be orientable as well so that the
integration of $B$ over $S_\a$ can be defined. We are able to find these
surfaces
only under particular conditions, suggesting that vertex operators exist only
under special circumstances of this sort.

Below, we will discuss some surfaces $S_\alpha $ associated with twisted spins
(which by definition have winding number $N$). When
it exists, the vertex operator can be constructed starting with an expression
like (\ref{4.39}) and regularising it following Section 4.2 {ii)}. We will
also write ${\cal C}$ and $S$ for ${\cal C}_\alpha $ and $S_\alpha $ when it
is convenient to do so.

The simplest construction of twisted spins is as follows. We start with a
ribbon with $L$ and $L'$ as borders and a flat surface interpolating them as
in Fig. 29 (a).
We then twist one end by $2\pi M$ and then identify the ends.
$M$ here is half integral or integral. The resultant configurations for
$M=\frac 12$ and 1 are shown in Figs. 29 (b) and (c).

Let us first consider Fig. 29(c). It shows an orientable surface $S$ with
vortices ${\cal C}$ and ${\cal C}'$ as borders. If the spin of ${\cal C}$
twists
by $+2\pi $ say, so that it has winding number $N=1$, then the spin of
${\cal C}'$ twists by $-2\pi$ and has $N=-1$.

Fig. 29(b) is a M\"{o}bius band. It is not orientable. For this reason, it
does not seem possible to create the vortices of this figure by a vertex
operator.


The method outlined here is capable of generalisations. One such would be to
first knot the ribbon, for example in the shape of a trefoil cut at a point, as
in Fig. 30(a).
The loose ends of the ribbon are then identified after $N$
twists. As $S$ becomes nonorientable if $N\in {\bf Z} +\frac 12$, $N$ here is
restricted to be an integer. $\pm N$ are then the winding numbers of the spins
of the vortices located at the borders $\cc $ and $\cc '$ indicated in
Fig. 30(b).

This example can be generalised by creating links and knots using several
ribbons,
with knots having twisted spins. Fig. 31 illustrates an example of this sort.
The
general idea here is the same as the one governing the passage from braids to
linked and unlinked knots \cite{knots}. The surface $S$ going into the
definition of the
vertex operator is the surface on ribbons. It would have disconnected
components
if the knot has several links.

There is  one further generalisation of this idea which gives the previous
constructions as special cases and also shows how to create new types of
states.
We recall that to create a state at $\cc_\a$, we first dig a hole $\ch_\a$
enclosing $\cc_\a$, which hole is eventually shrunk to $\cc_\a$. There is a
Fock space of states localised at $\partial \ch_\a$ with the Fock vacuum
$|0>$. The application of the vertex operator of a vortex involving a surface
with a boundary $\cc_\a$ gives the required vortex state.

Suppose now that there are holes $\ch_\a$ and $\ch_\b$ enclosing loops
$\cc_\a$ and $\cc_\b$ respectively. Let the cycles $Y_i$ introduced previously
be denoted by $Y_i(\a)$ and $Y_i(\b)$ when they are associated with $\ch_\a$
and $\ch_\b$. Let $Y(\r)$ [$\r=\a,~\b$] be a loop on $\partial \ch_\r$
which is homologous to $N_1(\r)Y_1(\r)+N_2(\r)Y_2(\r)$,
$N_1(\r)Y_1(\r)+N_2(\r)Y_2(\r)$ denoting the loop where $Y_1(\r)$ is
traversed $N_1(\r)$ times and $Y_2(\r)$ is traversed $N_2(\r)$ times.
Homologous curves for a given $\r$ are again defined by regarding them as
curves confined to $\partial \ch_\r$. Figure 32 exibits a typical $Y(\r)$
for a hole $\ch_\r$.

Assume that there exists a surface $S$ with one or more connected boundaries.
For a moment assume for specificity that there are two connected
boundaries $Y(\a)$ and $Y(\b)$.
If $|0>$ is the tensor product of the Fock vacua for $\partial \ch_\a$, we
can create a state $|Y(\a),Y(\b)>$ by applying the vertex operator involving
$S$ to $|0>$.

Now if for example $N_1(\a)=N_2(\a)=-N_1(\b)=-N_2(\b)=1$, we get the example
of Fig. 29(c). Simple generalisations of the construction leading to this
state will also yield all previous examples.

But we can also create new states now as the number of boundary components
need not be two and $N_i(\r )$ are not restricted to have
the values suitable for Fig.29(c). The surface $S_\a$ of Fig. 13 for example
is an instance where there is only one connected boundary. As another example,
suppose that $\ch_\a$ and
$\ch_\b$ are linked while $S$ has the property that
$N_1(\a)=0$, $N_2(\a)=1$, $N_1(\b)=-1$, $N_2(\b)=0$. Fig.33 shows how to
realise this situation.
In this case, when $\ch_\r$ finally shrunks to $\cc_\r$, there is a vortex
with flux associated with it at $\cc_\a$, but not at $\cc_\b$. Instead, we
have created winding number 1 spin excitations at $Y(\b)$, the definition of
this winding number being similar to its definition for a vortex with flux.

In this example, we can clearly deform $S$ so that $Y(\r)$ become any curve
homologous to the corresponding curve in Fig.33.

Note that $N_1(\r)$ measures the spin twist at $\cc_\r$. As for $N_2(\r)$,
suppose that for a particular $S$, $|N_2(\r)|$ is neither 0 or 1. Now,
$Y(\r)$ winds $N_2(\r)$ times around $\cc_\r$ and therefore after $\ch_\r$
shrinks to $\cc_\r$, the state describes a vortex at $\cc_\r$ with flux
$N_2(\r)\Phi$ if $\Phi_\a$ is $\Phi$ in the vertex operator defined by
(\ref{4.39}) or (\ref{4.57}).

The brief considerations presented in this Section show that we can create
several types of states by using suitable surfaces. But we will not pursue
their study further here.

Finally we point out that there is an operator which measures spin twist.
It can be constructed as follows. The state $|Y(\a),Y(\b)>$ discussed above
is given by
\be
|Y(\a),Y(\b)>=e^{-\imath \Phi \int_SB}|0> \label{7.1}
\ee
where $\Phi$ has been sunstituted for the $\Phi_\a$ of (\ref{4.39}). [More
precisely, we should define the state using the regularised version of the
vertex operator in (\ref{7.1})]. Let $\hat \cc_\r$ a curve just outside
the hole $\ch_\r$ and not touching $\partial \ch_\r$ obtained by deforming
$\cc_\r$ as shown in Fig.34.
Then we claim that
\be
\left[\int_{\hat \cc_\r}A, \int_{S}B \right]=\imath N_1(\r)\label{7.2}
\ee
and therefore that
\be
e^{\imath\Phi\int_SB}\,\,\left(\int_{\hat\cc_\r}A\right)
\,\,e^{-\imath\Phi\int_SB}=\Phi N_1(\r)\,. \label{7.3}
\ee
Since the expectation value of
\be
T(\r)=\frac {1}{\Phi} \int_{\hat \cc_\r}A \label{7.4}
\ee
in $|0>$ is zero, it follows that its expectation value in $|Y(\a),Y(\b)>$
measures the spin twist. For this reason, it is natural to regard $T(\r)$
as the spin twist operator for $\cc_\r$.

Before showing (\ref{7.2}), let us note that we can let $\hat \cc_\r$ tend
to $\cc_\r$ after $\ch_\r$ is shrunk to $\cc_\r$. Also, the mode
expansion of (\ref{7.4}) should be discussed, but we will not do so here.

As the first step in showing (\ref{7.2}), let us deform $\hat \cc_\r$ to a
curve $\tilde \cc_\r$ lying on $\partial \ch_\r$. $\tilde \cc_\r $ is a
cycle homologous to $Y_2(\r)$ and is shown in Figures 34 and 35.
Now it is well known that $\tilde \cc_\r$ will intersect $Y(\r)$ exactly
$N_1(\r)$ times if an intersection is counted as $+1$ or $-1$ according
to its orientation. Now the surface $S$ terminates at $Y(\r)$. Therefore,
as illustrated in Fig.36,
the loop $\hat \cc_\r$, obtained by lifting
$\tilde \cc_\r$ a little bit off $\partial \ch_\r$, will also intersect
$S$ exactly $N_1(\r)$ times. As each such intersection contributes an
$\imath$ or $-\imath$ to the left hand side of (\ref{7.2}) depending on its
orientation, the result (\ref{7.2}) is immediate.

If the expectation value of $T(\r)$ in $|Y(\a),Y(\b)>$ is not zero, then the
definition (\ref{7.4}) of $T(\r)$ suggests that $\hat \cc_\r$ encloses flux.
This implies in particular that a single unknotted vortex with zero flux
passing through its middle can not have spin twist.

\newpage

\centerline{ACKNOWLEDGEMENTS}

We have enjoyed many fruitful discussions with Peppe Bimonte, Elisa Ercolessi,
Kumar Gupta and Ajit Srivastava. This work was supported by the Department of
Energy and contract number DE-FG02-85ER40231. PTS also thanks CAPES (Brazil)
for partial support.

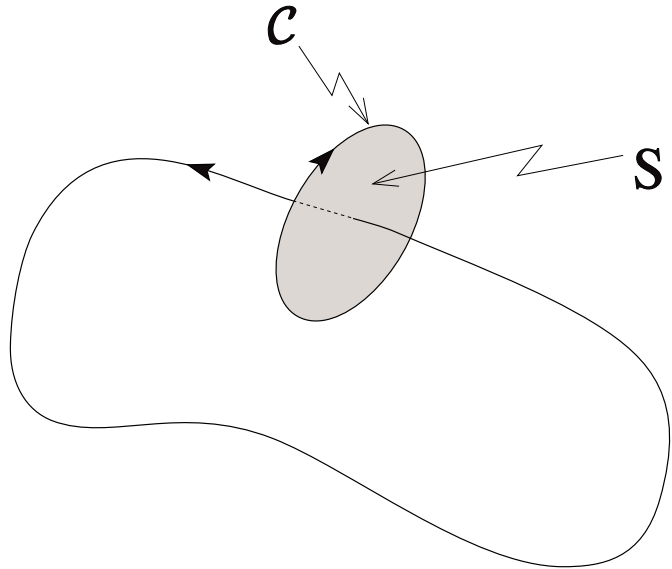
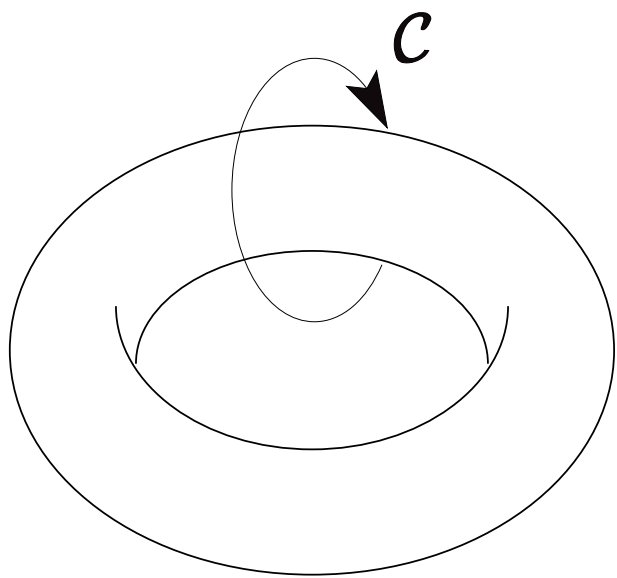
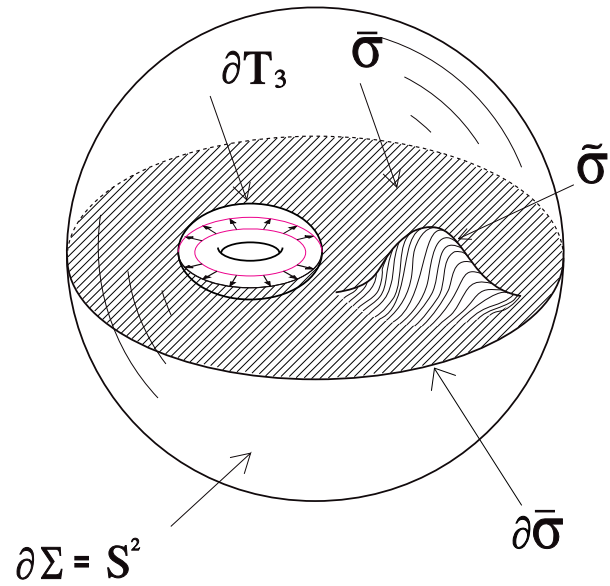
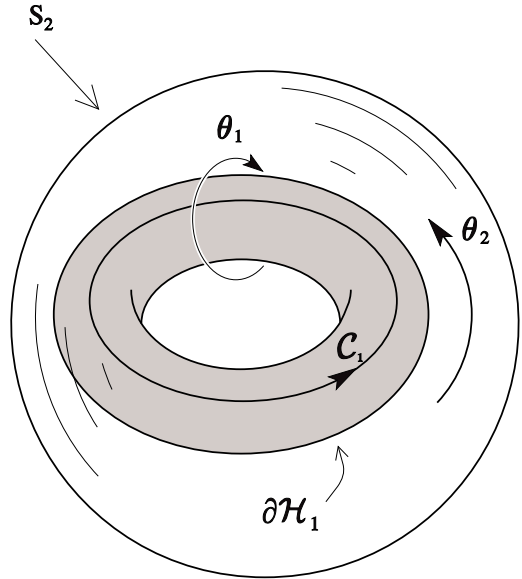
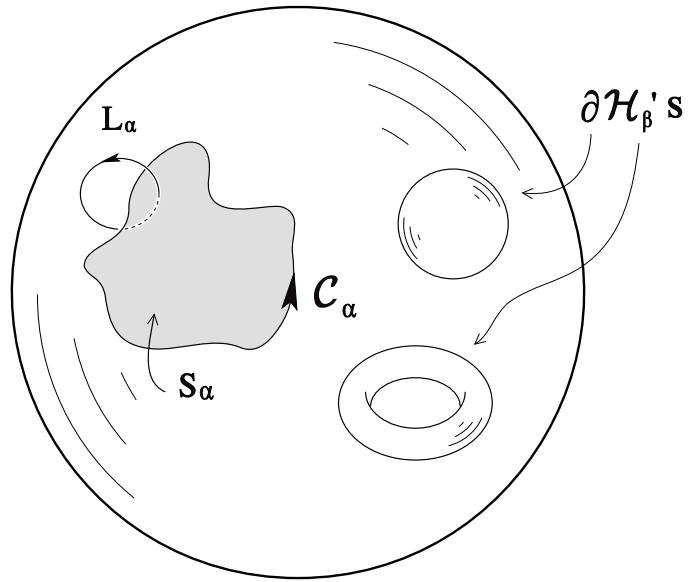
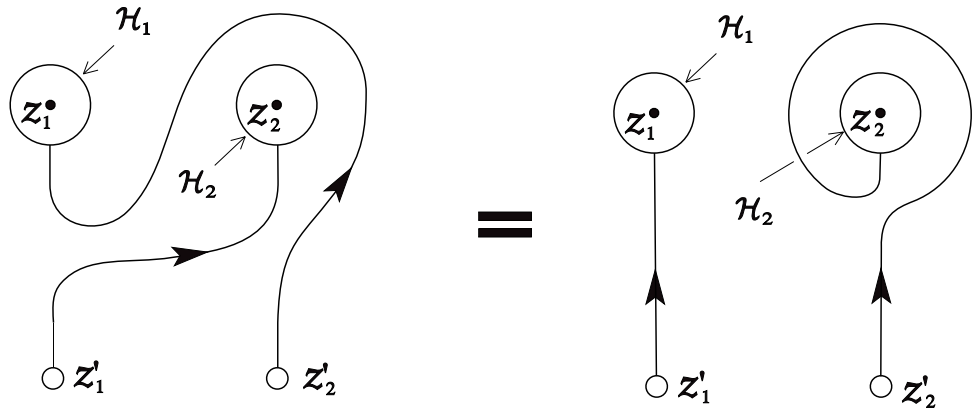
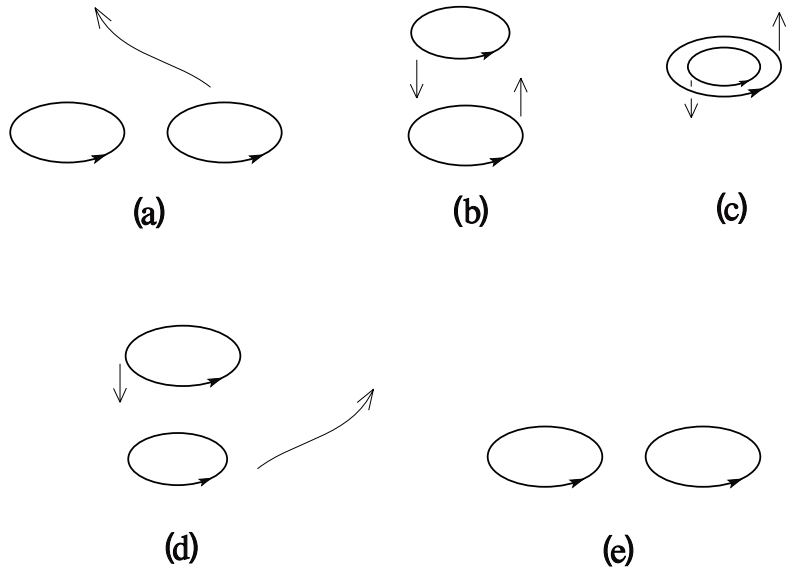
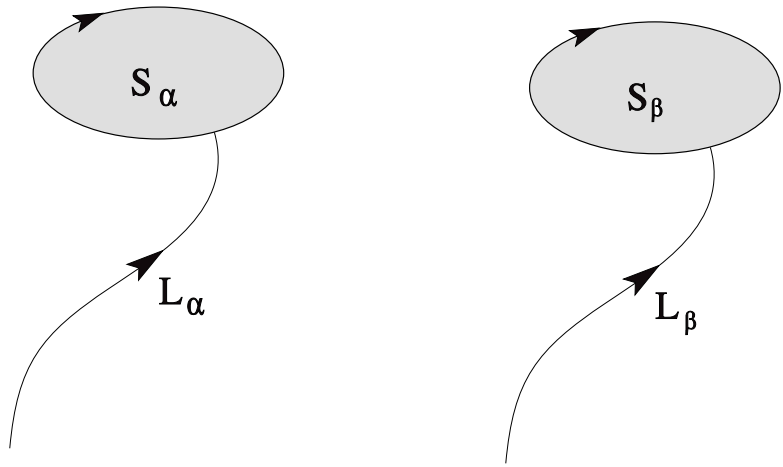
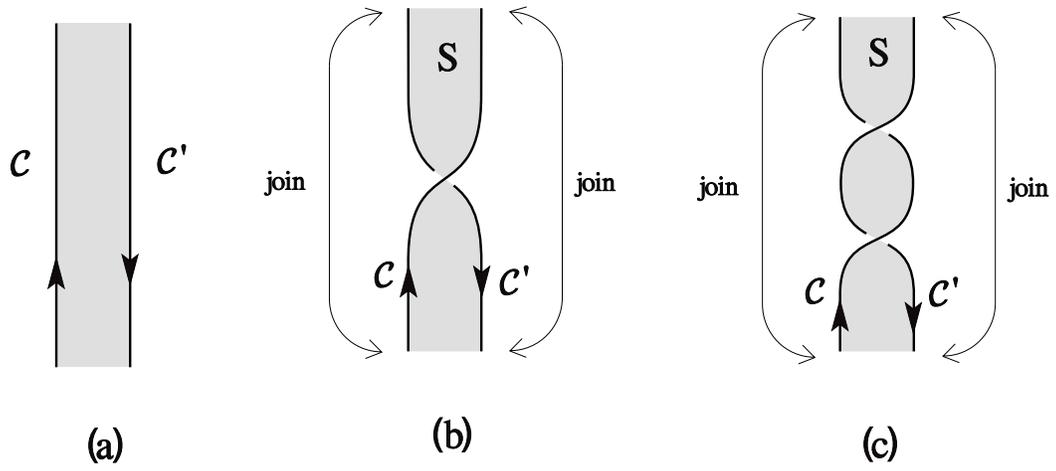
\end{document}